\documentclass[12pt]{article}
\usepackage{graphicx}
\usepackage{amssymb}
\usepackage{amsmath}
\usepackage{amsfonts}
\usepackage{anysize}
\marginsize{2.5cm}{2.5cm}{2.5cm}{3cm}
\date{}
\begin{document}


\title{\textbf{Generalized Statistical Mechanics at the Onset of Chaos}}
\author{Alberto Robledo \\ \\ {\footnotesize Instituto de F\' isica y Centro de Ciencias de la Complejidad} \\ {\footnotesize Universidad Nacional Aut\' onoma de M\' exico} \\ {\footnotesize \textbf{robledo@fisica.unam.mx}}}

\maketitle 

\abstract{Transitions to chaos in archetypal low-dimensional nonlinear maps offer real and precise model systems in which to assess proposed generalizations of statistical mechanics. The known association of chaotic dynamics with the structure of Boltzmann--Gibbs (BG) statistical mechanics has suggested the potential verification of these generalizations at the onset of chaos, when the only Lyapunov exponent vanishes and ergodic and mixing properties cease to hold. There are three well-known routes to chaos in these deterministic dissipative systems, period-doubling, quasi-periodicity and intermittency, which provide the setting in which to explore the limit of validity of the standard BG structure. It has been shown that there is a rich and intricate behavior for both the dynamics within and towards the attractors at the onset of chaos and that these two kinds of properties are linked via generalized statistical-mechanical expressions. Amongst the topics presented are: (i) permanently growing sensitivity fluctuations and their infinite family of generalized Pesin identities; (ii) the emergence of statistical-mechanical structures in the dynamics along the routes to chaos; (iii) dynamical hierarchies with modular organization; and (iv) limit distributions of sums of deterministic variables. The occurrence of generalized entropy properties in condensed-matter physical systems is illustrated by considering critical fluctuations, localization transition and glass formation. We complete our presentation with the description of the manifestations of the dynamics at the transitions to chaos in various kinds of complex systems, such as, frequency and size rank distributions and complex network images of time series. We discuss the results.} \\





\section{Introduction}

Chaotic dynamical systems, such as chaotic attractors in one-dimensional nonlinear iterated maps, accept a statistical-mechanical description with an entropy expression of the Boltzmann--Gibbs (BG) type \cite{beck1}. The chaotic attractors generated by these maps have ergodic and mixing properties, and not surprisingly, they can be described by a thermodynamic
formalism compatible with BG statistics \cite{beck1}. However, at the transition
to chaos, such as the period-doubling accumulation point, the so-called
Feigenbaum attractor, these two properties are lost, and this suggests the
possibility of exploring the limit of validity of the BG structure in a
precise, but simple enough, setting.

The attractors at the transitions to chaos in one-dimensional nonlinear maps
(that we shall refer to as critical attractors) have a vanishing ordinary
Lyapunov coefficient, $\lambda _{1}$, and the sensitivity to initial
conditions, $\xi _{t}$, for large iteration time $t$ ceases to obey
exponential behavior, exhibiting, instead, power-law or faster than
exponential growth behavior \cite{mori1, baldovin1, robledo1, robledo2, robledo3, baldovin1b}. As it is generally
suggested, the standard exponential divergence of trajectories in chaotic
attractors provides a mechanism to justify the assumption of irreversibility
in the BG statistical mechanics \cite{beck1}. In contrast, the onset of
chaos in (necessarily dissipative) one-dimensional maps imprints memory
preserving phase-space properties to its trajectories~\cite{mori1}, and we
consider them here with a view to assess if generalized entropy expressions
replace the usual BG expression. A simple type of attractor with $%
\lambda _{1}=0$ occurs at the transitions between periodic orbits, the
so-called pitchfork bifurcations along the period-doubling cascades \cite
{schuster1, hilborn1}. Another type, a critical attractor-repellor
pair, occurs at the tangent bifurcation \cite{schuster1, hilborn1},
a central feature for the intermittency route to chaos. Perhaps the most
interesting types of critical attractors are geometrically involved
(multifractal) sets of positions for which trajectories within them display
an ever fluctuating non-exponential sensitivity, $\xi _{t}$ \cite{mori1}.
Important examples are the onset of chaos via period doubling and via
quasiperiodicity, two other universal routes to chaos exhibited,
respectively, by the prototypical logistic and circle maps \cite{schuster1, hilborn1}.

There are two sets of properties associated with the attractors involved:
those of the dynamics inside the attractors and those of the dynamics
towards the attractors. These properties have been characterized in detail
for the Feigenbaum attractor; both the organization of trajectories and the
sensitivity to initial conditions are described in \cite{robledo1},
while the features of the rate of approach of an ensemble of trajectories to
this attractor appears, explained in \cite{robledo2}. The dynamics
inside quasiperiodic critical attractors has also been determined by
considering the so-called golden route to chaos in the circle map~\cite
{robledo3}. The patterns formed by trajectories and the sensitivity, $\xi
_{t}$, in this different case were found to share the general scheme
displayed by these quantities for the Feigenbaum point, but with somewhat more
involved features \cite{robledo3}. On the contrary, the critical
attractor-repellor set at the tangent bifurcation is a simpler finite set of
positions with periodic dynamics that can be reduced to a single period-one
attractor-repellor fixed point via functional composition, and the
interesting properties of trajectories and sensitivity, $\xi _{t}$, correspond
to the dynamics towards the attractor or away from the repellor \cite{baldovin1}.
The search for generalized entropy properties in the dynamics of critical
attractors has been two-fold. One focal point has been the determination of
Pesin-like identities associated with the sensitivity, $\xi _{t}$, within
multifractal critical attractors for which the entropy expression involved
is different from the BG expression. The second focal point is the
determination of the relationship between the dynamics towards and the
dynamics within multifractal critical attractors for which it has been found
that the expression linking the two types of dynamics is
statistical-mechanical in nature, one in which the configurational weights
and the thermodynamical potential expressions differ from the BG ordinary
exponential and logarithmic expressions. The simpler case of the tangent
bifurcation also involves generalized entropy expressions linked to the
trajectories and the sensitivity of the dynamics towards or away from the
attractor-repellor fixed point.

The generalized entropy expressions were found to be of the Tsallis or $q$
-statistical type \cite{tsallis1, tsallis2}, involving the ``$q$%
-deformed" logarithmic function $\ln _{q}y\equiv (y^{1-q}-1)/(1-q)$, the
functional inverse of the ``$q$-deformed" exponential function $\exp
_{q}(x)\equiv \lbrack 1-(q-1)x]^{-1/(q-1)}$.\ For the pitchfork and tangent
bifurcations, the trajectories and the sensitivity, $\xi _{t}$ (of their
functional-composition renormalization-group (RG) fixed-point maps), are
exactly expressible by $q$-exponentials, and this leads to generalized $q$%
-Lyapunov exponents $\lambda _{q}$ that are dependent on initial conditions 
\cite{baldovin1}. For multifractal critical attractors, the sensitivity, $\xi
_{t}$, was found to be expressible in terms of an infinite set of $q$%
-exponential functions, and from such $\xi _{t}$, one or several spectra of $%
q $-generalized Lyapunov coefficients $\lambda _{q}$ were determined. The $%
\lambda _{q}$ are dependent on the initial position, $x_{0}$, and each
spectrum can be examined by varying this position. The $\lambda _{q}$
satisfy a $q$-generalized identity $\lambda _{q}=K_{q}$ of the Pesin type 
\cite{hilborn1}, where $K_{q}$ is an entropy production rate based on $q$
-statistical-type entropy expressions $S_{q}$ \cite{robledo1, robledo3}. In all cases, the presence of dual $q$-indexes, $q$ and $2-q$, or $q$ and $1/q$, or a combination of them, have all well-defined dynamical meanings. If there is an entropy, $S_{q}$, or a coefficient, $\lambda _{q}$, in
the description of a critical attractor, the quantities, $S_{2-q}$ and $
\lambda _{2-q}$, complement that description \cite{robledo1, robledo3}
. The partition function that describes the approach of trajectories to the
Feigenbaum attractor contains an infinite set of $q$-exponential weights
(stemming from the infinite set of universal constants associated with the
attractor), and the resulting thermodynamic potential involves another $q$%
-index, which is the complicated outcome of all $q$-indexes involved in the
configurational terms \cite{robledo2}. The generalized entropy in this case
differs from the scheme contained in \cite{tsallis1, tsallis2}.

The manifestation of $q$-statistics in complex condensed-matter systems can
be explored via connections that have been established between the dynamics
of critical attractors and the dynamics taking place in, for example,
thermal systems under conditions when mixing and ergodic properties are not
easily fulfilled. A relationship between intermittency and critical
fluctuations suggested \cite{athens1, athens2} that the dynamics in
the proximity of a tangent bifurcation is analogous to the dynamics of
fluctuations of an equilibrium state with well-known scaling properties \cite{robcrit1, robcrit1b}. We examine this connection with special attention to several
unorthodox properties, such as, the extensivity of the $q$-entropy $S_{q}$
of fractal clusters of the order parameter and the anomalous (faster than
exponential) sensitivity to initial conditions. A second example is that of
the localization transition in electron transport in networks of scatterers. 
The use of a double Cayley tree model leads to an exact analogy
with a nonlinear map that features two tangent bifurcations. The
localization length is the inverse of the Lyapunov exponent, $\lambda _{1}$,
and when this vanishes, we obtain a precise description of the mobility edge
between insulating and conducting phases in terms of the $q$-generalized
exponent, $\lambda _{q}$ \cite{robloc1}. As a third example, we describe our 
finding \cite{robglass1, robglass1b, robglass1c} that the dynamics at the noise-perturbed
period-doubling onset of chaos is analogous to that observed in supercooled
liquids close to vitrification. We have demonstrated that four major
features of glassy dynamics in structural glass formers are displayed by
orbits with a vanishing Lyapunov coefficient. These are: two-step relaxation,
a relationship between relaxation time and configurational entropy, aging
scaling properties and evolution from a diffusive regime to arrest. The known
properties in control-parameter space of the noise-induced bifurcation gap
in the period-doubling cascade \cite{schuster1} play a central role in
determining the characteristics of dynamical relaxation at the chaos
threshold.

The manifestations of $q$-statistics in complex systems of a general or
interdisciplinary nature have been and are currently being exposed and
established. Two illustrations of general nature are provided: one
corresponds to dynamical hierarchies with modular organization and emergent
properties. We recall the demonstration that the dynamics towards
multifractal attractors as illustrated by the Feigenbaum point fulfills the main 
features of such dynamical hierarchies \cite{robledo4}. The other
general topic is that of limit distributions of sums of deterministic
variables, for which we describe and discuss the stationary distributions at
the period-doubling transition to chaos and how they transform to Gaussian
distributions for chaotic attractors. We consider both cases, that of an initial
trajectory inside the attractor and that of an ensemble of trajectories uniformly
distributed throughout all of phase space \cite{fuentes1, fuentes1b, fuentes2}.
We consider also two examples directed at more specific types of complex
systems. One case corresponds to the large class of systems described by
rank distributions. Their rank distributions are proved to be analogous to
the dynamics at or near the tangent bifurcation \cite{robledo5}. Another set
of studies correspond to the transcription of the time series formed by the
trajectories generated by the critical attractors into networks via an
appropriate algorithm known as the horizontal visibility method \cite{robtolo1}. Generalized entropies play an important role in understanding
these applications.

The structure of the rest of article is as follows: In Section 2, we provide
the basic description of the anomalous dynamics at critical attractors that
forms the core knowledge for the rest of the article. Of the three
routes to chaos in one-dimensional nonlinear maps, we refer only to the
intermittency and the period-doubling cases. For the sake of brevity, we do
not provide a description of the quasiperiodic critical attractors and give
only the relevant references. In Section 3, we extend our description to the
intricate dynamics towards the Feigenbaum attractor and refer to its
hierarchical structure with modular organization. In Section 4, we describe
the stationary distributions produced by sums of positions in the dynamics
associated with the Feigenbaum attractor and how they transform beyond the
transition to chaos. In Section 5, we present our applications to
condensed-matter complex systems, whereas in Section 6, we do so likewise for
general or interdisciplinary complex systems. In Section 7, we discuss our
results and make clarifying remarks.

\section{Critical Attractors in Unimodal Maps}

\subsection{Two Different Routes to Chaos in Unimodal Maps}

A unimodal map (a one-dimensional map with one extremum) contains infinite
families of critical attractors with vanishing Lyapunov exponent $\lambda
_{1}=0$ at which the ergodic and mixing properties breakdown \cite{note1}.
These are the tangent bifurcations that give rise to windows of periodic
trajectories within chaotic bands and the accumulation point(s) of the
pitchfork bifurcations, the so-called period-doubling onset of chaos \cite%
{schuster1, hilborn1} at which these periodic windows come to an
end. There are other attractors for which the Lyapunov exponent, $\lambda
_{1}$, diverges to minus infinity, where there is faster than exponential
convergence of orbits. These are the superstable attractors located between
successive pitchfork bifurcations. They are present at the initial
period-doubling cascade and at all the other cascades within periodic
windows with accumulation points that are replicas of the Feigenbaum
attractor.

The properties of the critical attractors are universal in the
renormalization-group (RG) sense, that is, all maps, $f(x)$, that lead to the
same fixed-point map, $f^{\ast }(x)$, under a repeated functional composition
and rescaling transformation share the same scaling properties. For unimodal
maps, this transformation takes the form $Rf(x)\equiv \alpha f(f(x/\alpha ))$%
, where $\alpha$ assumes a fixed value (positive or negative real number)
for each universality class and $f^{\ast }(x)$ is given by:
\begin{equation}
f^{\ast }(x)\equiv \lim_{n\rightarrow \infty }R^{(n)}f(x)=\lim_{n\rightarrow
\infty }\alpha ^{n}f^{(2^{n})}(x/\alpha ^{n}) \label{fixedpointmap1}
\end{equation}%
and satisfies:%
\begin{equation}
f^{\ast }(x)=\alpha f^{\ast }(f^{\ast }(x/\alpha )) \label{fixedpointmap2}
\end{equation}%
The universality of the static or geometrical properties of multifractal
critical attractors has been understood since long ago \cite{schuster1, hilborn1}. This is represented, for example, by the generalized dimensions,~$D_{q}$, or the spectrum, $f(\widetilde{\alpha })$, that characterize the
multifractal attractor at the period-doubling onset of chaos~\cite{beck1, hilborn1}. As we see below, the dynamical properties of critical
attractors, multifractal or not, also display universality, the entropic
index, $q$, in the sensitivity, $\xi _{t}$, and the Lyapunov spectra, $\lambda
_{q}$, is given in terms of the universal constant, $\alpha$. For the the
pitchfork and tangent bifurcations, the results are relatively
straightforward, but for the period-doubling accumulation point, the situation
is more complex. In the latter case, an infinite set of universal constants,
of which $\alpha$ is most prominent, is required. These constants are
associated with the discontinuities off the trajectory scaling function, $\sigma 
$, that measures the convergence of positions in the orbits of period $2^{n}$
as $n\rightarrow \infty$ to the Feigenbaum attractor \cite{schuster1}.

The sensitivity to initial conditions, $\xi _{t}$, is defined as:
\begin{equation}
\xi _{t}(x_{0})\equiv \lim_{\Delta x_{0}\rightarrow 0}(\Delta x_{t}/\Delta
x_{0}) 
\label{sensitivity}
\end{equation}%
where $\Delta x_{0}$ is the initial separation of two orbits and $\Delta
x_{t}$ that at time $t$. Notice, we do not write the customary limit, $t\gg 1$,
above. As we shall see when $\lambda _{1}$ vanishes, $\xi _{t}$ has the form 
\cite{baldovin1, baldovin2, robledo8}:
\begin{equation}
\xi _{t}(x_{0})=\exp _{q}[\lambda _{q}(x_{0})\ t]\equiv \lbrack
1-(q-1)\lambda _{q}(x_{0})\ t]^{-1/(q-1)} \label{sensitivity1}
\end{equation}%
that yields the standard exponential, $\xi _{t}$, with $\lambda _{1}$, when $
q\rightarrow 1$. In Equation (\ref{sensitivity1}), $q$ is the entropic index and $
\lambda _{q}$ is the $q$-generalized Lyapunov exponent; $\exp _{q}(x)\equiv
\lbrack 1-(q-1)x]^{-1/(q-1)}$ is the $q$-exponential function. The local
rate of entropy production, $K_{1}$, for chaotic attractors is given by $
K_{1}t=S_{1}(t)-S_{1}(0)$ and $S_{1}=-\sum_{i}p_{i}\ln p_{i}$ with $p_{i}$,
the trajectories' distribution. For chaotic attractors, the identity $K_{1}=$ $\lambda _{1}>0$ holds \cite{hilborn1}.
Notice that this is not the Pesin identity that, instead of $K_{1}$, employs
the Kolmogorov--Sinai entropy, $\mathcal{K}_{1}$ \cite{robledo8}. For critical
attractors, the identity still holds trivially, but a generalized form:
\begin{equation}
K_{q}=\lambda _{q}
\label{q=pesin1}
\end{equation}%
appears where the rate of $q$-entropy production, $K_{q}$, is defined via: 
\begin{equation}
K_{q}t=S_{q}(t)-S_{q}(0)
\label{q-entropyrate1}
\end{equation}
and where:
\begin{equation}
S_{q}\equiv \sum_{i}p_{i}\ln _{q}\left( \frac{1}{p_{i}}\right) =\frac{1-\sum_{i}^{W}p_{i}^{q}}{q-1} 
\label{tsallisentropy1}
\end{equation}%
is the Tsallis entropy (Recall that $\ln _{q}y\equiv (y^{1-q}-1)/(1-q)$ is
the functional inverse of $\exp _{q}(y)$).

We take as a starting point and framework for the study of fixed-point map
properties the prototypical logistic map, or its generalization to
the non-linearity of order $\zeta >1$:
\begin{equation}
f_{\mu }(x)=1-\mu \left\vert x\right\vert ^{\zeta },\;-1\leq x\leq 1,0\leq
\mu \leq 2 
\label{z-logistic1}
\end{equation}%
where $x$ is the phase space variable, $\mu$ the control parameter and $
\zeta =2$ corresponds to the familiar logistic map. Our results relate to
the anomalous $\xi _{t}$ and its associated spectrum, $\lambda _{q}$, for the
above-mentioned critical attractors that are involved in the two routes to
chaos exhibited by unimodal maps, the intermittency and the period doubling
routes. For the Feigenbaum attractor, we describe the relationship of $\xi
_{t}$ and $\lambda _{q}$ with Mori's $q$-phase transitions \cite{mori1, robledo1}.

\subsection{Dynamics at the Tangent and Pitchfork Bifurcations}

The exact fixed-point map solution of the RG Equation (\ref{fixedpointmap2}) for
the tangent bifurcations, known since long ago \cite{schuster1, hilborn1}, has been shown to describe the dynamics of iterates in
the neighborhood of this attractor \cite{robledo9, robledo9b, baldovin1}, as well. Furthermore,
a straightforward extension of this approach was shown to apply to the
pitchfork bifurcations \cite{robledo9, robledo9b, baldovin1}. We recall that
period-doubling and intermittency transitions are based on the pitchfork and
the tangent bifurcations, respectively, and that at these attractors, the
ordinary Lyapunov coefficient $\lambda _{1}=0$. The sensitivity, $\xi _{t}$,
can be determined analytically and its relation with the rate of entropy
production examined \cite{robledo9, robledo9b}. The fixed-point expressions have the
specific form that corresponds to the temporal evolution suggested by $q$
-statistics. In \cite{robledo9, robledo9b, baldovin1} is contained the derivation
of the $q$-Lyapunov coefficients, $\lambda _{q}$, and the description of the
different possible types of sensitivity, $\xi _{t}$.

For the transition to periodicity of order $n$ in the $\zeta $-logistic map,
the composition, $f_{\mu }^{(n)}$, is first considered. In the neighborhood of
one of the $n$ points tangent to the line with unit slope, one obtains: 
\begin{equation}
f^{(n)}(x)=x+u\left\vert x\right\vert ^{z}+o(\left\vert x\right\vert ^{z})
\label{n-thf1}
\end{equation}
where $u$ is the expansion coefficient. At the tangent bifurcations, one has $
z=2$ and$\ u>0$, whereas for the pitchfork bifurcations, one has instead $z=3$
, because $d^{2}f_{\mu }^{(2^{k})}/dx^{2}=0$ at these transitions, and $u<0$
is now the coefficient associated with $d^{3}f_{\mu }^{(2^{k})}/dx^{3}<0$.

\begin{figure}
\centering
\begin{tabular}{cc}
\includegraphics[width=0.4\textwidth]{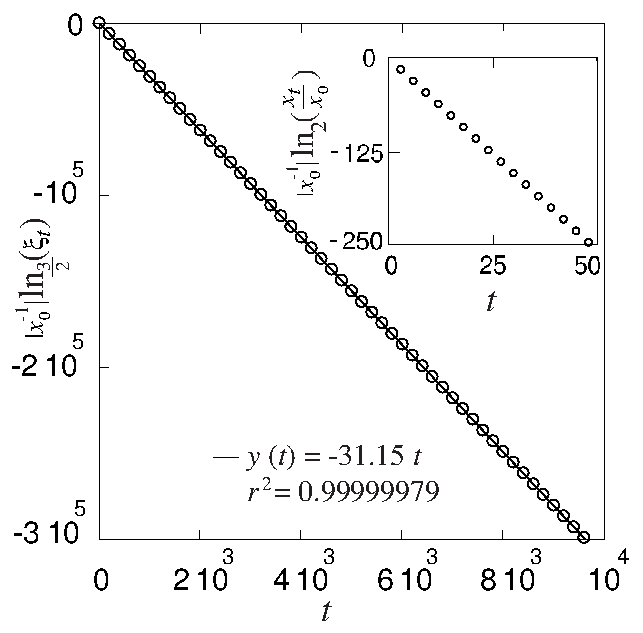} & 
\includegraphics[width=0.4\textwidth]{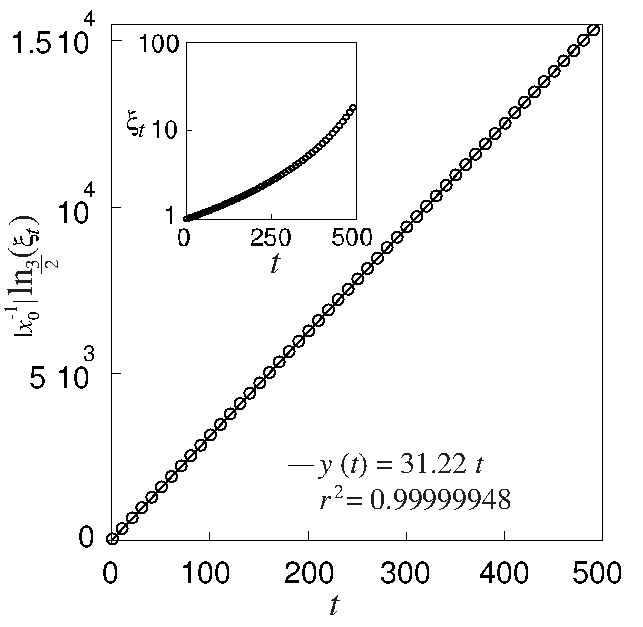} \\
(a) & (b) \\
\end{tabular}
\caption{Numerical corroboration of Eq. (\ref{sensitivity2}) for the tangent bifurcation. Panel (a) corresponds to the left side of the first tangent bifurcation for $\protect
\zeta =2$. Circles represent $|x_{0}^{-1}|\ln _{\frac{3}{2}}(\protect\xi 
_{t})$ for the iterates of $f^{(3)}$ and $x_{0}\sim -10^{-4}$. Panel (b) corresponds to the right side of the first tangent bifurcation for $
\protect\zeta =2$. Circles represent $|x_{0}^{-1}|\ln _{\frac{3}{2}}(\protect
\xi_{t})$ for the iterates of $f^{(3)}$ and $x_{0}\sim +10^{-4}$. See text for description.} 
\label{Fig._4}
\end{figure}

\begin{figure}
\centering
\includegraphics[width=0.4\textwidth]{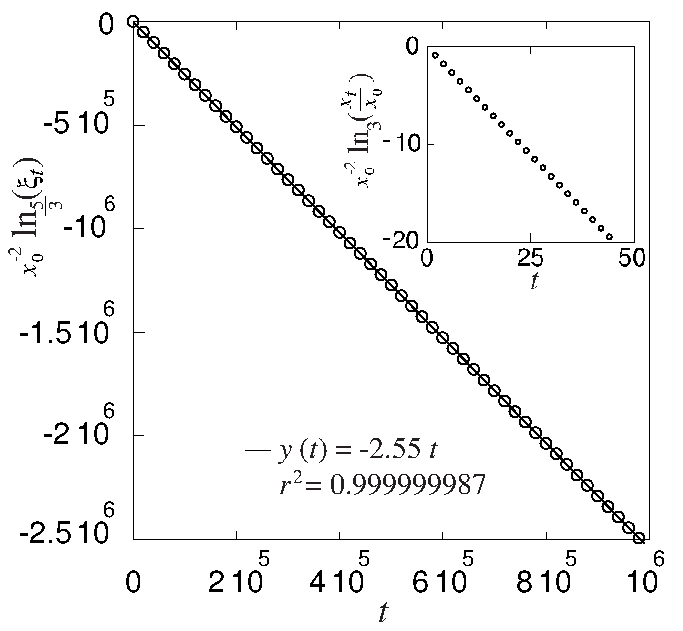}
\caption{Numerical corroboration of Eq. (\ref{sensitivity2}) for the first pitchfork bifurcation with $\protect\zeta =1.75$. Circles
represent $x_{0}^{-2}\ln _{\frac{5}{3}}(\protect\xi _{t})$ for the iterates
of $f^{(2)}$ and $x_{0}\sim +10^{-3}$. See text for description.} 
\label{Fig._5}
\end{figure}

The RG fixed-point map $x^{\prime }=f^{\ast }(x)$ associated with maps of
the form in Equation (\ref{n-thf1}) was found \cite{hu1} to be:
\begin{equation}
x^{\prime }=x\exp _{z}(ux^{z-1})=x[1-(z-1)ux^{z-1}]^{-1/(z-1)}
\label{fixed1}
\end{equation}%
as it satisfies $f^{\ast }(f^{\ast }(x))=\alpha ^{-1}f^{\ast }(\alpha x)$
with $\alpha =2^{1/(z-1)}$ and has a power-series expansion in $x$ that
coincides with Equation (\ref{n-thf1}) in the two lowest-order terms. (above $%
x^{z-1}\equiv \left\vert x\right\vert ^{z-1}\mathrm{sgn}(x)$). The long time
dynamics is readily derived from the static solution Equation (\ref{fixed1}); one
obtains:
\begin{equation}
\xi _{t}(x_{0})=[1-(z-1)ax_{0}^{z-1}t]^{-z/(z-1)},\;u=at
\label{sensitivity2}
\end{equation}%
and so, $q=2-z^{-1}$ and $\lambda _{q}(x_{0})=zax_{0}^{z-1}$ \cite{robledo1, baldovin1}. When $q>1$ the left-hand side ($x<0$) of the tangent
bifurcation map, Equation (\ref{n-thf1}) exhibits a weak insensitivity to
initial conditions, {\it i.e.}, power-law convergence of orbits. However at the
right-hand side ($x>0$) of the bifurcation, the argument of the $q$%
-exponential becomes positive, and this results in a ``super-strong''
sensitivity to initial conditions, {\it i.e.}, a sensitivity that grows faster than
exponential \cite{baldovin1}. For the tangent bifurcation, one has $z=2$ in $%
q=2-z^{-1}$, and so, $q=3/2$. For the pitchfork bifurcation, one has $z=3$ in $%
q=2-z^{-1}$, and one obtains $q=5/3$. Notably, these specific results for the
index $q$ are valid for all $\zeta >1$ and, therefore, define the existence of
only two universality classes for unimodal maps, one for the tangent and the
other one for the pitchfork bifurcations \cite{baldovin1}. See Figures \ref{Fig._4} and \ref{Fig._5} for numerical corroboration of the above. In Figure \ref{Fig._4}(a) the full line is a linear regression, whose slope, $-31.15$, should be compared with the
exact expression for the generalized Lyapunov exponent, which gives $\protect
\lambda_{q}=31.216...\;$ Inset: $|x_{0}^{-1}|\ln _{2}(x_{t}/x_{0})$
, for $t=3m\;\;\;(m=1,2,...)$ and $x_{0}\sim -10^{-6}$. A linear regression
gives, in this case, a slope of $-5.16 $, to be compared with the exact result 
$u/3=5.203...\;$ In Figure \ref{Fig._4}(b) the full
line is a linear regression with a slope equal to $31.22$. Inset: Log-linear
plot of $\protect\xi_{t}$ evidencing a super-exponential growth. In Figure \ref{Fig._5} the full line is a linear regression,
whose slope, $-2.55$, should be compared with the analytical expression for
the generalized Lyapunov exponent, which gives $\protect\lambda%
_{q}=-2.547...\;$ Inset: $x_{0}^{-2}\ln _{3}(x_{t}/x_{0})$, for $
t=2m\;\;\;(m=1,2,...)$ and $x_{0}\sim +10^{-3}$. A linear regression gives,
in this case, a slope of $-0.44$, to be compared with the analytical result $
u/2=-0.424...\;$.


There is an interesting scaling property displayed by $\xi _{t}$ in Equation (\ref
{sensitivity2}) similar to the scaling property known as \textit{aging} in
systems close to glass formation. This property is observed in two-time
functions (e.g., time correlations) for which there is no time translation
invariance, but scaling is observed in terms of a time ratio variable, $
t/t_{w}$, where $t_{w}$ is a ``waiting time'' assigned to the time interval
for the preparation or hold of the system before time evolution is observed
through time $t$. This property can be seen immediately in $\xi _{t}$ if one
assigns a waiting time, $t_{w}$, to the initial position, $x_{0}$, as $t_{w}=$ $
x_{0}^{1-z}$. Equation~(\ref{sensitivity2}) reads now:
\begin{equation}
\xi _{t,t_{w}}=[1-(z-1)at/t_{w}]^{-z/(z-1)} \label{sensitivity3}
\end{equation}%
The sensitivity for this critical attractor is dependent on the initial
position, $x_{0}$, or, equivalently, on its waiting time, $t_{w}$, the closer $
x_{0}$ is to the point of tangency, the longer $t_{w}$, but the sensitivity of
all trajectories fall on the same $q$-exponential curve when plotted against 
$t/t_{w}$. Aging has also been observed for the properties of the map in Equation (
\ref{n-thf1}), but in a different context \cite{barkai1}.

Notice that our treatment of the tangent bifurcation differs from other
studies of intermittency transitions \cite{gaspard1} in that there is no
feedback mechanism of iterates to indefinitely repeat the action of $
f^{(n)}(x)$ or of its associated fixed-point map, $f^{\ast }(x)$. Therefore,
impeded or incomplete mixing in phase space (a small interval neighborhood
around $x=0$) arises from the special ``tangency'' shape of the map at the
pitchfork and tangent transitions that produces monotonic trajectories. This
has the effect of confining or expelling trajectories, causing anomalous
phase-space sampling, in contrast to the thorough coverage in generic states
with $\lambda _{1}>0$. By construction, the dynamics at the intermittency
transitions describe a purely $q$-exponential regime.

\subsection{Dynamics within the Period-Doubling Accumulation Point} \label{SECdynamicsperdoubaccpoint}

The dynamics at the Feigenbaum attractor has been analyzed recently \cite
{robledo1,baldovin2,baldovin3}. For the $\zeta $-logistic
map, this attractor is located at $\mu _{\infty }$, the accumulation point of
the control parameter values for the pitchfork bifurcations, $\mu _{n}$, $%
n=1,2,...$, which is also that for the superstable orbits, $\overline{\mu }%
_{n} $, $n=1,2,...$. By taking as initial condition $x_{0}=0$ at $\mu
_{\infty }$, or, equivalently, $x_{1}=1$, it is found that the resulting
orbit, a superstable orbit of period $2^{\infty }$, consists of trajectories
made of intertwined power laws that asymptotically reproduce the entire
period-doubling cascade that occurs for $\mu <\mu _{\infty }$. This orbit
captures the properties of the superstable orbits that precedes it. Here,
again, the Lyapunov coefficient, $\lambda _{1}$, vanishes (although the
attractor is also the limit of a sequence of supercycles with $\lambda
_{1}\rightarrow -\infty $), and in its place, there appears a spectrum of $q$-Lyapunov coefficients, $\lambda _{q}^{(k)}$ (the $k$ index is defined below). The dynamics within this
attractor was originally studied in \cite{politi1, mori1}, and
our interest has been to examine its properties in relation with the
expressions of the Tsallis statistics. We found that the sensitivity to
initial conditions has precisely the form of a set of interlaced $q$-exponentials, of which we determine the $q$-indexes and the associated $%
\lambda _{q}^{(k)}$. As mentioned, the appearance of a specific value for
the $q$ index (and actually, also, that for its conjugate value $Q=2-q$) turns
out to be due to the occurrence of Mori's ``$q$-phase transitions'' \cite%
{mori1} between ``local attractor structures'' at $\mu _{\infty }$.
Furthermore, it has also been shown \cite{baldovin3, robledo1} that
the dynamical and entropic properties at $\mu _{\infty }$ are naturally
linked through the $q$-exponential and $q$-logarithmic expressions,
respectively, for the sensitivity to initial conditions, $\xi _{t}$, and for
the entropy, $S_{q}$, in the rate of entropy production, $K_{q}^{(k)}$. We have
analytically corroborated the equality $\lambda _{q}^{(k)}=$ $K_{q}^{(k)}$.
Our results support the validity of the $q$-generalized (local $t$) Pesin
identity for critical attractors in low-dimensional maps.

\begin{figure} 
\begin{center}
\includegraphics[width=.45\textwidth, angle=270]{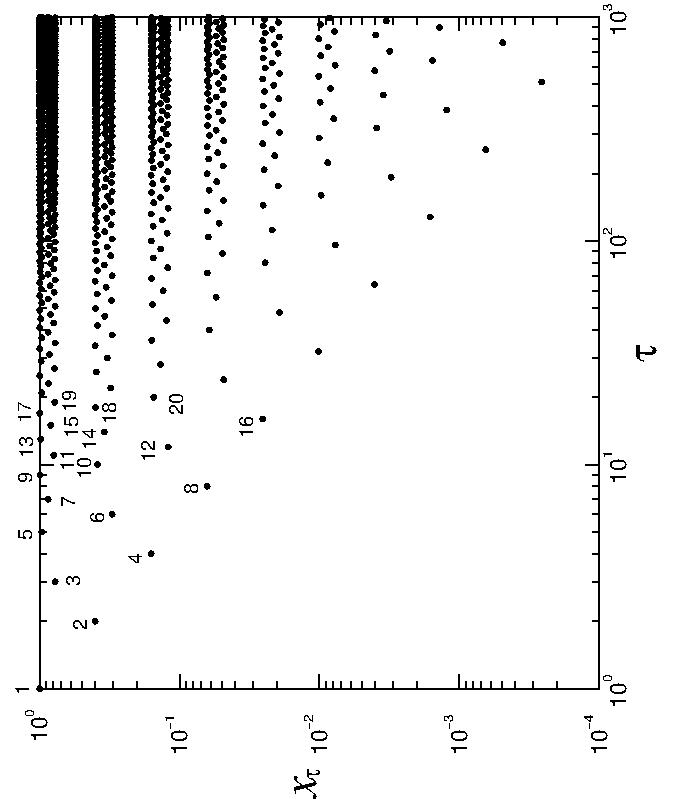}
\end{center}
\caption{Absolute values of positions in logarithmic scales of the first $%
1000$ iterations $\protect\tau$ for a trajectory of the logistic map at the
onset of chaos $\protect\mu_{\infty}$ with initial condition $x_{0}=0$. The
numbers correspond to iteration times. The power-law decay of the time
subsequences described in the text can be clearly appreciated.}
\label{Fig._6}
\end{figure}

More specifically, the absolute values for the positions $x_{\tau }$ of the
trajectory with $x_{t=0}=0$ at time-shifted $\tau =t+1$ have a structure
consisting of subsequences with a common power-law decay of the form $\tau
^{-1/1-q}$ with: 
\begin{equation}
q=1-\ln 2/(\zeta -1)\ln \alpha (\zeta )
\label{q-index1}
\end{equation}%
where $\alpha (\zeta )$ is the Feigenbaum universal constant for
nonlinearity $\zeta >1$ that measures the period-doubling amplification of
iterate positions \cite{baldovin2}. That is, the Feigenbaum attractor can be
decomposed into position subsequences generated by the time subsequences $%
\tau =(2k+1)2^{n}$, each obtained by proceeding through $n=0,1,2,...$ for a
fixed value of $k=0,1,2,...$. See Figure \ref{Fig._6}. The $k=0$ subsequence
can be written as $x_{t}=\exp _{2-q}(-\lambda _{q}^{(0)}t)$ with $\lambda
_{q}^{(0)}=(\zeta -1)\ln \alpha (\zeta )/\ln 2$. These properties follow
from the use of $x_{0}=0$ in the scaling relation \cite{baldovin2}: 

\begin{equation}
x_{\tau }\equiv \left\vert g^{^{(\tau )}}(x_{0})\right\vert =\tau
^{-1/1-q}\left\vert g(\tau ^{1/1-q}x_{0})\right\vert \label{trajectory1}
\end{equation}

where $g(x)$ is the Feigenbaum fixed-point map \cite{schuster1, hilborn1}.


\noindent The sensitivity associated with trajectories with other starting points, $
x_{0}\neq 0$, within the multifractal attractor (but located within either
its most sparse or most crowded regions) can be determined similarly with
the use of the time subsequences $\tau =(2k+1)2^{n}$. One obtains:
\begin{equation}
\lambda _{q}^{(k)}=\frac{(\zeta -1)\ln \alpha (\zeta )}{(2k+1)\ln 2}>0,\
k=0,1,2,... 
\label{q-lyapunov1a}
\end{equation}%
for the positive branch of the Lyapunov spectrum, when the trajectories
start at the most crowded ($x_{\tau =0}=1$) and finish at the most sparse ($%
x_{\tau =2^{n}}=0$) region of the attractor. By inverting the situation, we
obtain:
\begin{equation}
\lambda _{Q}^{(k)}=-\frac{2(\zeta -1)\ln \alpha (\zeta )}{(2k+1)\ln 2}<0,\
k=0,1,2,... 
\label{q-lyapunov1b}
\end{equation}%
for the negative branch of $\lambda _{q}^{(k)}$, {\it i.e.}, starting at the most
sparse ($x_{\tau =0}=0$) and finishing at the most crowded ($x_{\tau
=2^{n}+1}=1$) region of the attractor. Notice that $Q=2-q$ as $\exp
_{Q}(y)=1/\exp _{q}(-y)$. For the case $\zeta =2$, see \cite{baldovin2,baldovin3}; for general $\zeta >1$ see \cite{robledo1}, where,
also, different and more direct derivations are presented. Therefore, when
considering these two dominant families of orbits, all the $q$-Lyapunov
coefficients appear associated with only two specific values of the Tsallis
index, $q$ and $2-q$, with $q$ given by Equation (\ref{q-index1}). In Figure \ref%
{Fig._7}, we show the $q$-logarithm of $\xi _{t}~(x_{0}=1)$ \textit{vs}. $t$ for the $%
k=0$ time subsequence $\tau =2^{n}$ when $\zeta =2$ and $q=1-\ln 2/\ln
\alpha (2)=0.2445...$ and $\lambda _{q}^{(0)}=\ln \alpha (2)/\ln 2=1.3236...$.


\begin{figure} 
\begin{center}
\includegraphics[width=.4\textwidth]{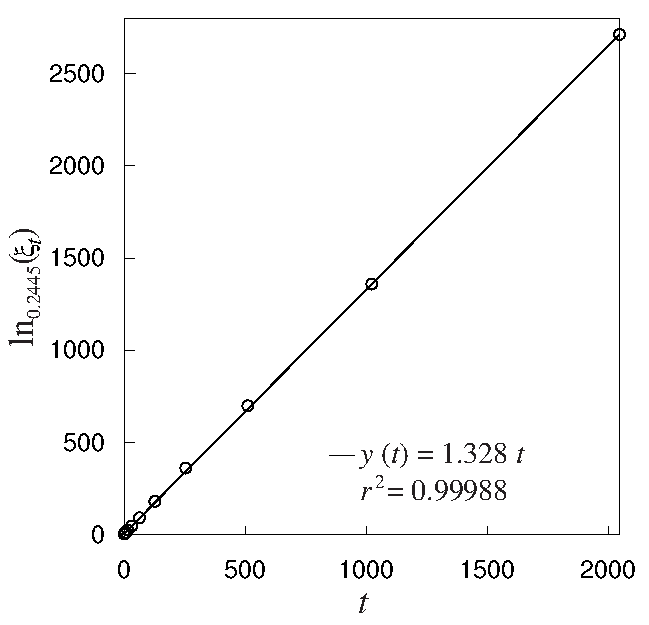}
\end{center}
\caption{The $q$-logarithm of sensitivity to initial conditions $\protect%
\xi _{t}$ \textit{vs}. $t$, with $q=1-\ln 2/\ln \protect\alpha =0.2445...$ and
initial conditions $x_{0}=0$ and $x_{0}=\protect\delta \simeq 10^{-8}$
(circles). The full line is the linear regression, $y(t)$. As required, the
numerical results reproduce a straight line with a slope very close to $%
\protect\lambda _{q}=\ln \protect\alpha /\ln 2=1.3236...\;$}
\label{Fig._7}
\end{figure}

\begin{figure} 
\begin{center}
\includegraphics[width=0.4\textwidth]{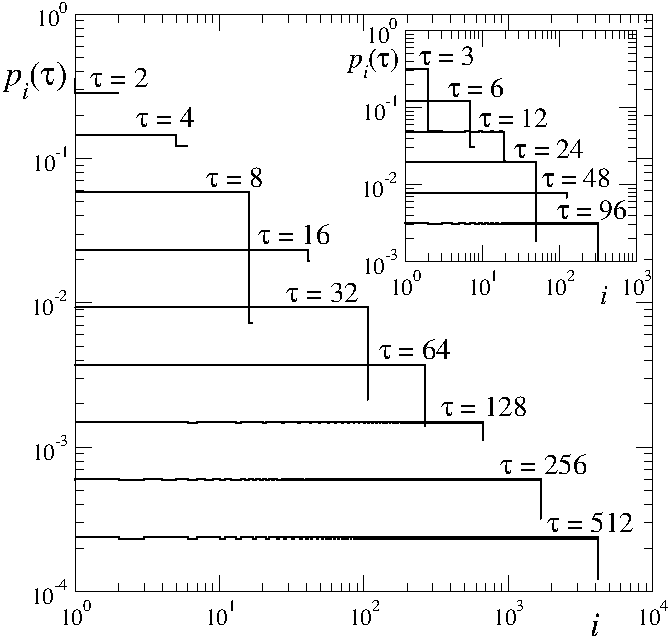}
\end{center}
\caption{{Time evolution, in logarithmic scales, of a
distribution, $p_{i}(\protect\tau )$, of trajectories at $\protect\mu %
_{\infty}$. Initial positions are contained within a cell adjacent to $x=1$,
and $i$ is the relative number of cells. Iteration time is shown for the
first two subsequences ($k=0,1$). }}
\label{Fig._8}
\end{figure}

\begin{figure} 
\begin{center}
\includegraphics[width=.4\textwidth]{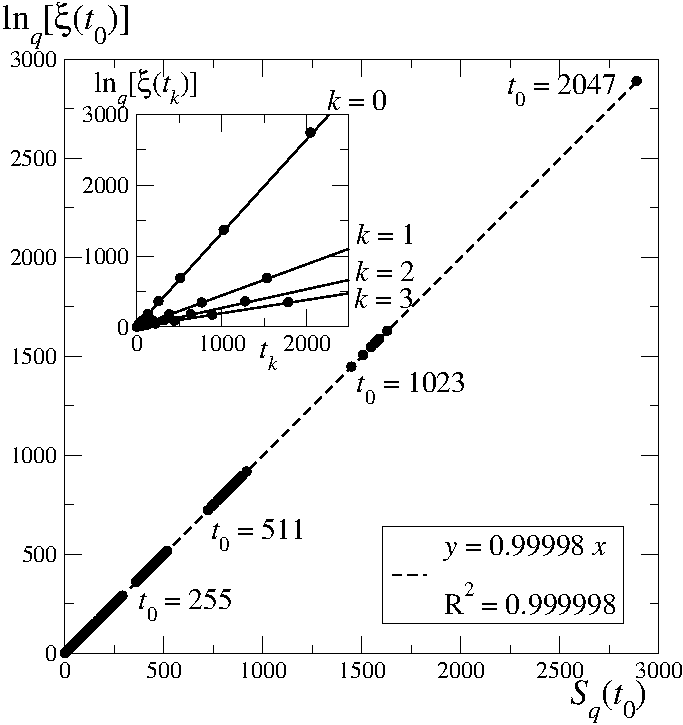}
\end{center}
\caption{Numerical corroboration (full circles) of the
generalized Pesin identity $K_{q}^{(k)}=\protect\lambda _{q}^{(k)}$ at $%
\protect\mu _{\infty }$. On the vertical axis, we plot the $q$-logarithm of $%
\protect\xi _{t_{k}}$ (equal to $\protect\lambda _{q}^{(k)}t)$ and in the
horizontal axis $S_{q}$ (equal to $K_{q}^{(k)}t$). In both cases, $q=1-\ln
2/\ln \protect\alpha =0.2445...$ The dashed line is a linear fit. In the
inset, the full lines are from analytical results. }
\label{Fig._9}
\end{figure}

Ensembles of trajectories with starting points close to $x_{\tau =0}=1$
expand in such a way that a uniform distribution of initial conditions
remains uniform for all later times, $t\leq T$, where $T$ marks the crossover
to an asymptotic regime. As a consequence of this, the identity of the rate
of entropy production, $K_{q}^{(k)}$, with $\lambda _{q}^{(k)}$ was
established \cite{baldovin3}. See Figures \ref{Fig._8} and \ref{Fig._9}. A
similar reasoning can be generalized to other starting positions \cite%
{robledo1}.


Notably, the appearance of a specific value for the $q$ index (and actually,
also, that for its conjugate value $Q=2-q$) turns out \cite{robledo1} to be
due to the occurrence of Mori's ``$q$-phase transitions'' \cite{mori1} between
``local attractor structures'' at $\mu _{\infty}$. To see this in more detail, we
observe that the sensitivity, $\xi _{t}(x_{0})$, can be obtained \cite
{robledo1} from:
\begin{equation}
\xi _{t}(m)\simeq \left\vert \frac{\sigma _{n}(m-1)}{\sigma _{n}(m)}
\right\vert ^{n},\;t=2^{n}-1,\;n\ 
\label{sensitivity4}
\end{equation}%
where $\sigma _{n}(m)=d_{n+1,m}/d_{n,m}$ and where $d_{n,m}$ are the
diameters that measure adjacent position distances that form the
period-doubling cascade sequence \cite{schuster1}. Above, the choices, $%
\Delta x_{0}=d_{n,m}$ and $\Delta x_{t}=d_{n,m+t}$, $t=2^{n}-1$, have been
made for the initial and the final separation of the trajectories,
respectively. In the large $n$ limit, $\sigma _{n}(m)$ develops
discontinuities at each rational of the form $m/2^{n+1}$ \cite{schuster1},
and according to the expression above for $\xi _{t}(m)$, the sensitivity is
determined by these discontinuities. For each discontinuity of $\sigma
_{n}(m)$, the sensitivity can be written in the forms \cite{robledo1}:
\begin{equation}
\xi _{t}=\exp _{q}[\lambda _{q}t],\;\lambda _{q}>0 \label{sensitivity5a}
\end{equation}%
and:%
\begin{equation}
\xi _{t}=\exp _{2-q}[\lambda _{2-q}t],\;\lambda _{2-q}<0,
\label{sensitivity5b}
\end{equation}%
where $q$ and the spectra, $\lambda _{q}$ and $\lambda _{2-q}$, depend on the
parameters that describe the discontinuity \cite{robledo1}. This result
reflects the multi-region nature of the multifractal attractor and the
memory retention of these regions in the dynamics. The pair of $q$
-exponentials correspond to a departing position in one region and arrival
at a different region and \textit{vice versa}; the trajectories expand in one sense
and contract in the other. The largest discontinuity of $\sigma _{n}(m)$ at $
m=0$ is associated with trajectories that start and finish at the most crowded
($x\simeq 1$) and the most sparse ($x\simeq 0$) regions of the attractor. In
this case, one obtains, again, Equation (\ref{q-lyapunov1a}), the positive branch of
the Lyapunov spectrum, when the trajectories start at $x\simeq 1$ and finish
at $x\simeq 0$. By inverting the situation, one obtains Equation (\ref
{q-lyapunov1b}), the negative branch of the Lyapunov spectrum. Therefore, when
considering these two dominant families of orbits, all the $q$-Lyapunov
coefficients appear associated with only two specific values of the Tsallis
index, $q$ and $Q=2-q$, with $q$ given by Equation (\ref{q-index1}).

\begin{figure} 
\begin{center}
\includegraphics[width=.72\textwidth]{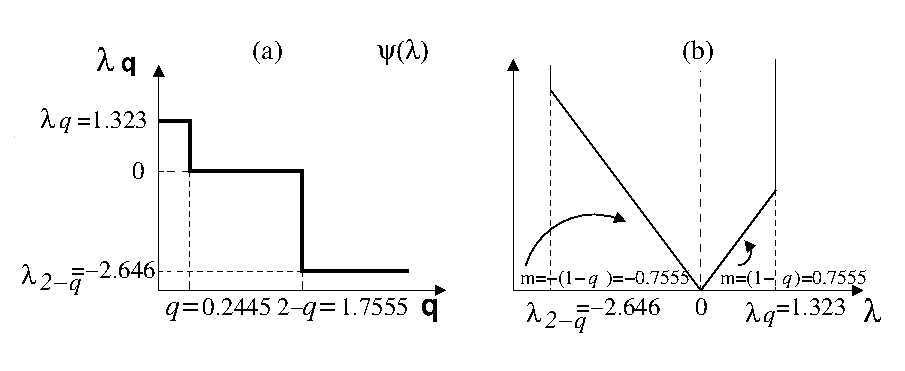}
\end{center}
\caption{(\textbf{a}) The Lyapunov coefficient function, $\protect\lambda (q)$, at the
chaos threshold at $\protect\mu _{\infty}$; and (\textbf{b}) the spectrum, $\protect\psi (
\protect\lambda ) $. See the text for a description.}
\label{Fig._10}
\end{figure}

As a function of the running variable $-\infty <\mathsf{q}<\infty$, the $q$
-Lyapunov coefficients become a function, $\lambda (\mathsf{q})$, with two
steps located at $\mathsf{q}=q=1-\ln 2/(\zeta -1)\ln \alpha (\zeta )$ and $
\mathsf{q}=Q=2-q$. See Figure~\ref{Fig._10}. In this manner, contact can be
established with the formalism developed by Mori and coworkers \cite{mori1}
and the $q$-phase transition obtained numerically in \cite{politi1, mori2}. The step function for $\lambda (\mathsf{q})$ can be integrated
to obtain the \textit{spectrum of local coefficients}, $\phi (\mathsf{q})$ ($
\lambda (\mathsf{q})\equiv d\phi /d\lambda (\mathsf{q})$), and its Legendre
transform, $\psi (\lambda )$ ($\equiv \phi -(1-\mathsf{q})\lambda $), the
dynamic counterparts of the Renyi dimensions, $D(\mathsf{q})$, and the
spectrum of local dimensions, $f(\widetilde{\alpha })$, that characterize the
geometry of the attractor. The result for $\psi (\lambda )$ is:
\begin{equation}
\psi (\lambda )=\left\{ 
\begin{array}{l}
(1-Q)\lambda ,\ \lambda _{Q}^{(0)}<\lambda <0 \\ 
(1-q)\lambda ,\ 0<\lambda <\lambda _{q}^{(0)}%
\end{array}%
\right. \label{psispectrum1}
\end{equation}%
As with ordinary thermal first order phase transitions, a $q$-phase transition
is indicated by a section of linear slope $1-q$ in the spectrum (free
energy), $\psi (\lambda )$, a discontinuity at $q$ in the Lyapunov function
(order parameter), $\lambda (\mathsf{q})$, and a divergence at $q$ in the
variance (susceptibility), $v(\mathsf{q})$. For the onset of chaos at $\mu
_{\infty }(\zeta =2)$, a $q$-phase transition was determined numerically \cite
{politi1, mori1, mori2}. According to $\psi (\lambda )$ above,
we obtain a conjugate pair of $q$-phase transitions that correspond to
trajectories linking two regions of the attractor, the most crowded and most
sparse. See Figure \ref{Fig._10}. Details appear in \cite{robledo1}. See
\cite{robledo3} for the derivation of the analog expressions for $\xi
_{t}$, $\lambda _{q}^{(k)}$, $K_{q}^{(k)}$, $\lambda (\mathsf{q})$ and $\psi
(\lambda )$ associated with the dynamics at the quasiperiodic critical
attractor via the golden ratio route to chaos in the circle map.

\section{Dynamics towards the Feigenbaum Attractor}
\vspace{-12pt}
\subsection{Diameters, Repellors and Gap Formation}

The knowledge of the dynamics towards a particular family of periodic
attractors, the so-called superstable attractors \cite{schuster1, hilborn1}, facilitates the understanding of the rate of approach of
trajectories to the Feigenbaum attractor, located at $\mu =\mu _{\infty }$,
and highlights the source of the discrete scale invariant property of this
rate \cite{robledo2}. The family of trajectories associated with these
attractors (also called supercycles) of periods $2^{N}$, $N=1,2,3,...$,
are located along the bifurcation forks. The positions (or phases) of the $%
2^{N}$-attractor are given by $x_{j}=f_{\overline{\mu }_{N}}^{(j)}(0)$, $%
j=1,2,\ldots ,2^{N}$. Notice that infinitely many other sequences of
superstable attractors appear at the period-doubling cascades within the
windows of periodic attractors for values of $\mu >$ $\mu _{\infty }$.
Associated with the $2^{N}$-attractor at $\mu =\overline{\mu }_{N}$, there is
a $(2^{N}-1)$-repellor consisting of $2^{N}-1$ positions $y_{k}$, $%
k=0,1,2,\ldots ,2^{N}-1$. These positions are the unstable solutions, $%
\left\vert df_{\overline{\mu }_{N}}^{(2^{n-1})}(y)/dy\right\vert >1$, of $%
y=f_{\overline{\mu }_{N}}^{(2^{n-1})}(y)$, $n=1,2,\ldots ,N$. The first, $%
n=1$, originates at the initial period-doubling bifurcation; the next two, $%
n=2$, start at the second bifurcation, and so on, with the last group of $%
2^{N-1}$, $n=N$, setting out from the $N-1$ bifurcation. The diameters, $%
d_{N,m}$, are defined as $d_{N,m}\equiv x_{m}-f_{\overline{\mu }%
_{N}}^{(2^{N-1})}(x_{m})$.

\begin{figure} 
\begin{center}
\includegraphics[width=.6\textwidth]{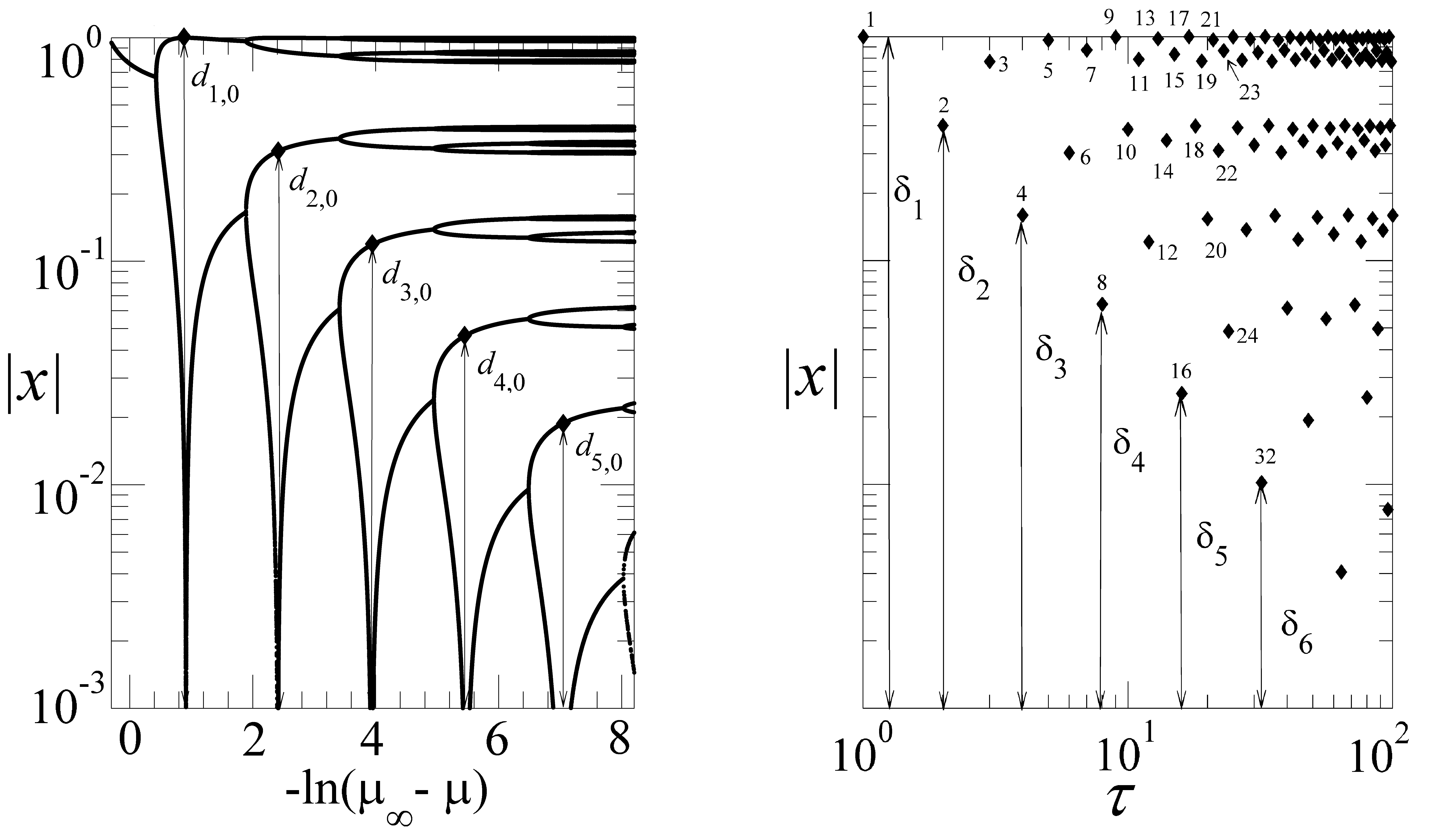}
\end{center}
\caption{(\textbf{Left panel}) Absolute value of attractor positions
for the logistic map, ${\protect\small f}_{\protect\mu }{\protect\small (x)}$%
{\protect\small\ , in logarithmic scale, as a function of }${\protect\small -}%
\ln {\protect\small (\protect\mu) }_{\infty }{\protect\small -\protect\mu}$%
{\protect\small. (\textbf{Right panel}) Absolute value of trajectory positions for }$%
{\protect\small f}_{\protect\mu }{\protect\small (x)}${\protect\small \ at }$%
{\protect\small \protect\mu }_{\infty }${\protect\small \ with initial
condition }${\protect\small x}_{0}{\protect\small =0}${\protect\small \ in
logarithmic scale, as a function of the logarithm of time} ${\protect\small 
\protect\tau }$; $\protect\tau$ is also shown by the
numbers close to the diamonds. The arrows indicate the equivalence between
the diameters ${\protect\small d}_{N,0}${\protect\small \ in the left
panel and position differences }${\protect\small \protect\delta }_{n}$%
{\protect\small \ with respect to }${\protect\small x}_{0}{\protect\small =0}
${\protect\small \ in the right panel.}}
\label{trajmap}
\end{figure}

Central to our understanding of the dynamical properties of unimodal maps is
the following in-depth property: Time evolution at $\mu _{\infty }$ from $%
\tau =0$ up to $\tau \rightarrow \infty$ traces the period-doubling cascade
progression from $\mu =0$ up to $\mu _{\infty }$. There is an underlying
quantitative relationship between the two developments. Specifically, the
trajectory inside the Feigenbaum attractor with initial condition $x_{0}=0$,
the $2^{\infty }$-supercycle orbit, takes positions $x_{\tau }$, such that
the distances between appropriate pairs of them reproduce the diameters, $%
d_{N,m}$, defined from the supercycle orbits with $\overline{\mu }_{N}<\mu
_{\infty }$. See Figure \ref{trajmap}, where the absolute value of
positions and logarithmic scales are used to illustrate the equivalence.
This property has been basic in obtaining rigorous results for the
sensitivity to initial conditions for the Feigenbaum attractor \cite
{robledo1} and for the dynamics of approach to this attractor \cite
{robledo2}. Other families of periodic attractors share most of the
properties of supercycles. Below, we consider explicitly the case of a map
with a quadratic maximum, but the results are easily extended to general
nonlinearity $\zeta >1$.

\begin{figure} 
\centering
\includegraphics[width=0.7\textwidth]{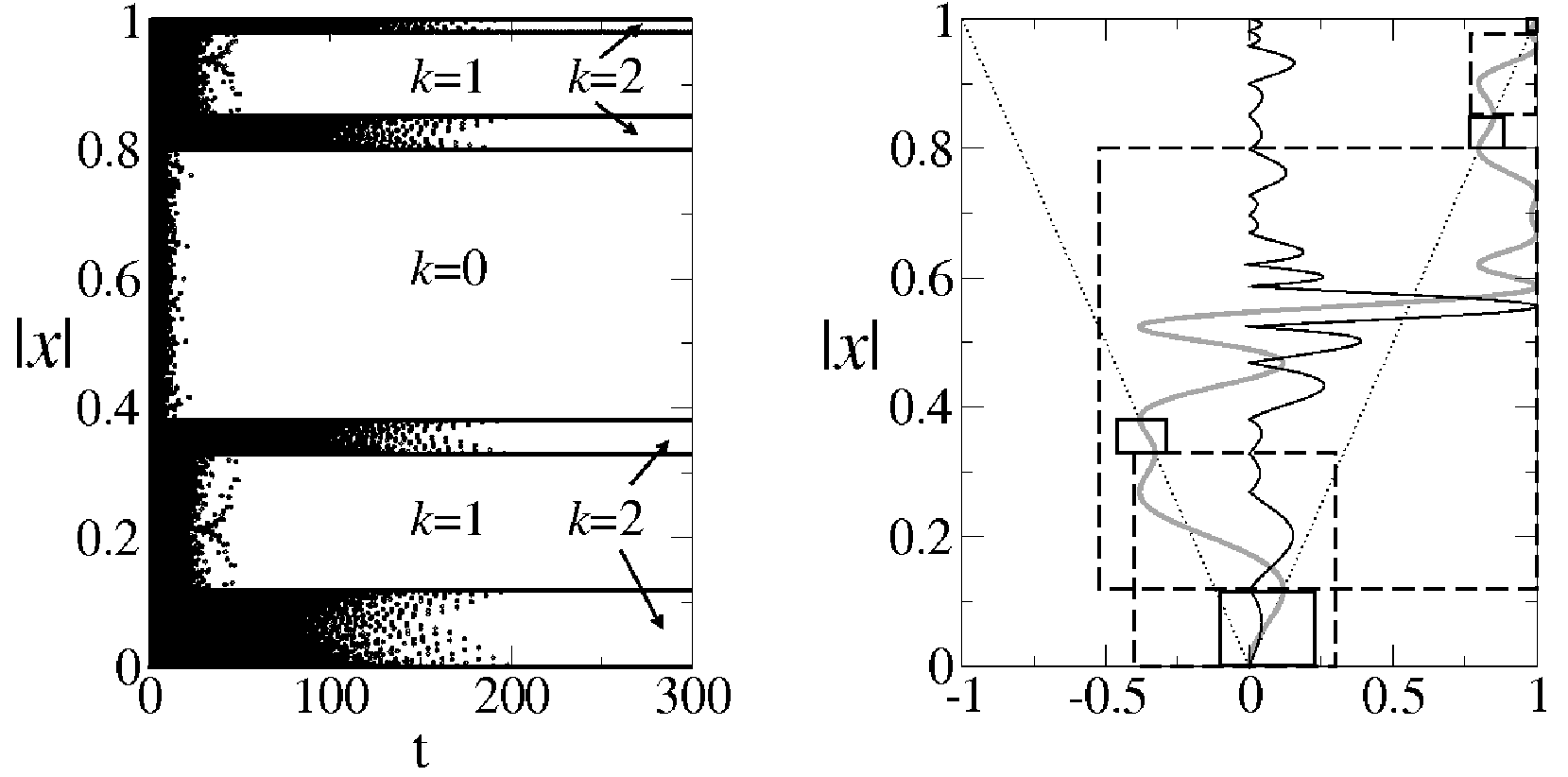}
\caption{\small Phase-space gap formation for $\mu = \mu_3$. (\textbf{Left panel}) Time evolution of a uniform ensemble of $10^4$ trajectories as a function of $\vert x \vert$ (black areas and open circles). The values of the index $k$ label the order of the gap set. (\textbf{Right panel}) Rotated plots of $f_{\overline{\mu}_{3}}^{(8)}(x)$ and as guides for the identification of attractor and repellor positions.} 
\label{fig:f15pre08}
\end{figure}

The organization of the total set of trajectories as generated by all
possible initial conditions as they flow towards a period, $2^{N}$, attractor
has been determined in detail \cite{robledo2, robledo10}. It was
found that the paths taken by the full set of trajectories in their way to
the supercycle attractors (or to their complementary repellors) are
exceptionally structured. The dynamics associated with families of
trajectories always displays a characteristically concerted order in which
positions are visited, and this is reflected in the dynamics of the
supercycles of periods $2^{N}$ via the successive formation of gaps in phase
space (the interval $-1\leq x\leq 1$) that finally give rise to the
attractor and repellor multifractal sets. See Figure \ref{fig:f15pre08}. To observe explicitly this process,
an ensemble of initial conditions, $x_{0}$, distributed uniformly across phase
space, was considered, and their positions were recorded at subsequent times 
\cite{robledo2, robledo10}. This set of gaps develops in time,
beginning with the largest one associated with the first repellor position,
then followed by a set of two gaps associated with the next two repellor
positions; next, a set of four gaps associated with the four next repellor
positions, and so forth. The gaps that form consecutively all have the same
width in the logarithmic scales (see Figures 14--17 in \cite{robledo2}), and
therefore, their actual widths decrease as a power law, the same power law
followed, for instance, by the position sequence $x_{\tau }=\alpha ^{-N}$, $%
\tau =2^{N}$, $N=0,1,2,...$ for the trajectory inside the attractor
starting at $x_{0}=0$ (and where $\alpha \simeq 2.50291$ is the absolute
value of Feigenbaum's universal constant for $\zeta =2$). The locations of
this specific family of consecutive gaps (the largest gaps for each value of 
$k$) advance monotonically toward the sparsest region of the multifractal
attractor located at $x=0$. See \cite{robledo2, robledo10} for
more details.


\subsection{Sums of Diameters as Partition Functions}

\begin{figure} 
\centering
\includegraphics[width=0.6\textwidth]{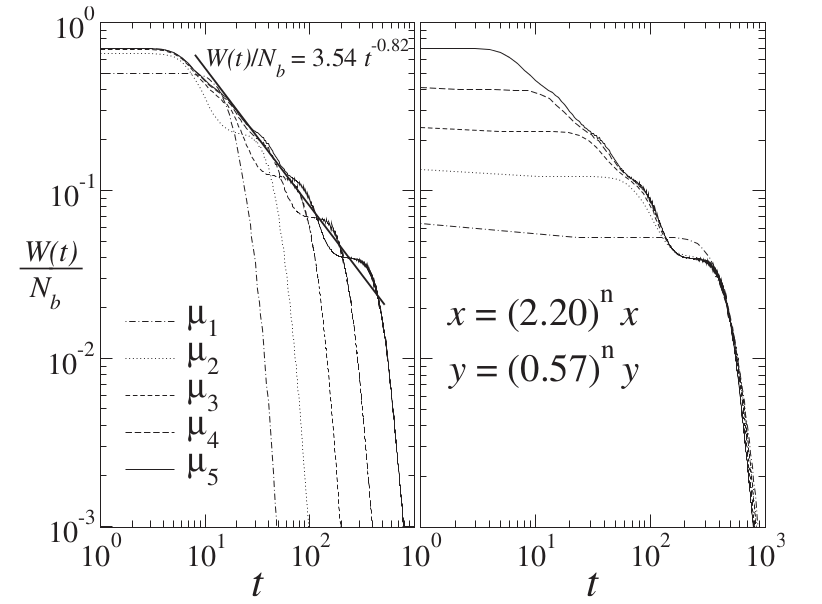}
\caption{
\small \textbf{(Left panel)} Rate $W_t$, divided by the number of boxes,
$N_b$, employed, of an approach to the attractor for the supercycles of
periods $2^N, N = 1, 2, 3, 4$ and 5 on logarithmic scales. The expression shown corresponds to the power law decay of the developing
logarithmic oscillations. Right panel: Superposition of the five
curves for $W_t$ in the left panel via $n$-times repeated rescaling
factors shown for the horizontal $x$ and vertical $y$ axes.}
\label{fig:f19pre08}
\end{figure}

The rate of convergence, $W_{t}$, of an ensemble of trajectories towards any
attractor/repellor pair along the period-doubling cascade is a convenient
single-time quantity that has a straightforward definition and is practical
to implement numerically. A partition of phase space is made of $N_{b}$
equally-sized boxes or bins, and a uniform distribution of $N_{c}$ initial
conditions is placed along the interval, $-1\leq x\leq 1$. The ratio, $
N_{c}/N_{b}$, can be adjusted to achieve optimal numerical results \cite
{robledo2}. The quantity of interest is the number of boxes, $W_{t}$, that
contain trajectories at time $t$. This rate has been determined for the
supercycles $\overline{\mu }_{N}$, $N=1,2,3,...$ and its accumulation point 
$\mu _{\infty }$ \cite{robledo2}. See Figure \ref{fig:f19pre08}, where 
$W_{t}$ is shown in logarithmic scales for the first five supercycles of
periods $2^{1}$ to $2^{5}$, where we can observe the following features: In
all cases, $W_{t}$ shows a similar initial and nearly constant plateau $%
W_{t}\simeq \Delta $, $1\leq t\leq t_{0}$, $t_{0}=O(1)$, and a final
well-defined decay to zero. The jump $\Delta$ at $\mu _{\infty }$ is $%
\Delta =(1+\alpha ^{-1})/2\simeq 0.69977$, due to the fact that
all initial conditions out of the interval ($-\alpha ^{-1}$,$1$) take a
value inside this interval after the first iteration. As can be observed
in the left panel of Figure \ref{fig:f19pre08}, the time position of the
final decay grows approximately proportionally to the period $2^{N}$ of the
supercycle. There is an intermediate slow decay of $W_{t}$ that develops as $%
N$ increases with duration, also just about proportional to $2^{N}$. For the
shortest period, $2^{1}$, there is no intermediate feature in $W_{t}$; this
appears first for period $2^{2}$ as a single dip and expands with one
undulation every time $N$ increases by one unit. The expanding intermediate
regime exhibits the development of a power-law decay with logarithmic
oscillations (characteristic of discrete scale invariance). In the limit, $%
N\rightarrow \infty $, the rate takes the form $W_{t}\simeq \Delta h(\ln \tau
/\ln 2)\tau ^{-\phi}$, $\tau =t-t_{0}$, where $h(x)$ is a periodic function
with $h(1)=1$ and $\phi \simeq 0.8001$ \cite{robledo2}.

The rate, $W_{t}$, at the values of time for period doubling, $\tau =2^{n}$, $%
n=1,2,3,...<N$, can be obtained quantitatively from the supercycle diameters 
$d_{n,m}$. Specifically:

\begin{equation}
Z_{\tau }\equiv \frac{W_{t}}{\Delta }=\sum_{m=0}^{2^{n-1}-1}d_{n,m},\ \tau
=t-t_{0}=2^{n-1},\ n=1,2,3,...<N 
\label{partition1}
\end{equation}

\noindent Equation (\ref{partition1}) expresses the numerical procedure followed
in \cite{grassberger1} to evaluate the exponent, $\phi$, but it also suggests a
statistical-mechanical structure if $Z_{\tau }$ is identified as a partition
function, where the diameters, $d_{n,m}$, play the role of configurational
terms \cite{robledo2}. The diameters, $d_{N,m}$, scale with $N$ for $m$ fixed
as $d_{N,m}\simeq \alpha _{y}^{-(N-1)}$, $N$ large, where the $\alpha _{y}$
are universal constants obtained from the finite discontinuities of
Feigenbaum's trajectory scaling function $\sigma (y)=\lim_{N\rightarrow
\infty }(d_{N,m+1}/d_{N,m})$, $y=\lim_{N\rightarrow \infty }(m/2^{N})$ \cite
{schuster1, robledo2}. The largest two discontinuities of $\sigma (y)$
correspond to the sparsest and denser regions of the multifractal attractor
at $\mu _{\infty }$, for which we have, respectively, $d_{N,0}\simeq \alpha
^{-(N-1)}$ and $d_{N,1}\simeq \alpha ^{-2(N-1)}$ ($d_{1,0}=1$). The
diameters, $d_{N,m}$, can be rewritten exactly as $q$-exponentials via use of the identity $
A^{-(N+1)}\equiv (1+\beta)^{-\ln A/\ln 2}$, $\beta =2^{N-1}-1$. That
is, $d_{N,m}\simeq \exp _{q_{y}}(-\beta\epsilon_{y} )$, where, $%
q_{y}=1-\epsilon_{y}^{-1}$, $\epsilon_{y}=-\ln \alpha _{y}/\ln 2$, and $\beta
=\tau -1=2^{N-1}-1$. Similarly, $Z_{\tau }\simeq \tau ^{-\phi}$ can be
expressed as $Z_{\tau }\simeq \exp _{\mathcal{Q}}(-\beta\phi )$, where $
\mathcal{Q}=1-\phi^{-1}$. Therefore, taking the above into account in Equation (\ref
{partition1}), we have:

\begin{equation}
\exp _{\mathcal{Q}}(-\beta\phi )\simeq \sum_{y}\exp _{q_{y}}(-\beta\epsilon
_{y} )
\label{partition2}
\end{equation}

\noindent Equation (\ref{partition2}) resembles a basic statistical-mechanical
expression with the exception that $q$-deformed exponential weights appear
in place of ordinary exponential weights (that are recovered when $\mathcal{Q
}=q_{y}=1$). We emphasize that the left- and right-hand sides in Equations (\ref
{partition1}) and (\ref{partition2}) are given by quantities that describe
the dynamics of approach and within the attractor, respectively. See 
\cite{robledo2, robdiaz1} for related details.

\subsection{Dynamical Hierarchies with Modular Organization}

We point out \cite{robledo4} that it is possible to generate hierarchical
systems with dynamical organization from a formal starting point in the form
of a simple closed-form nonlinear dynamical system. Furthermore, that nonlinear
dynamics tuned at the transition to chaos can lead to an emergent property.
We illustrate this standpoint by appropriate interpretation to the
properties of the dynamics towards the Feigenbaum attractor we are
discussing.

\subsubsection{Preimage Structure and Flow of Trajectories towards the
Attractor}

\begin{figure} 
\centering
\includegraphics[width=0.45\textwidth]{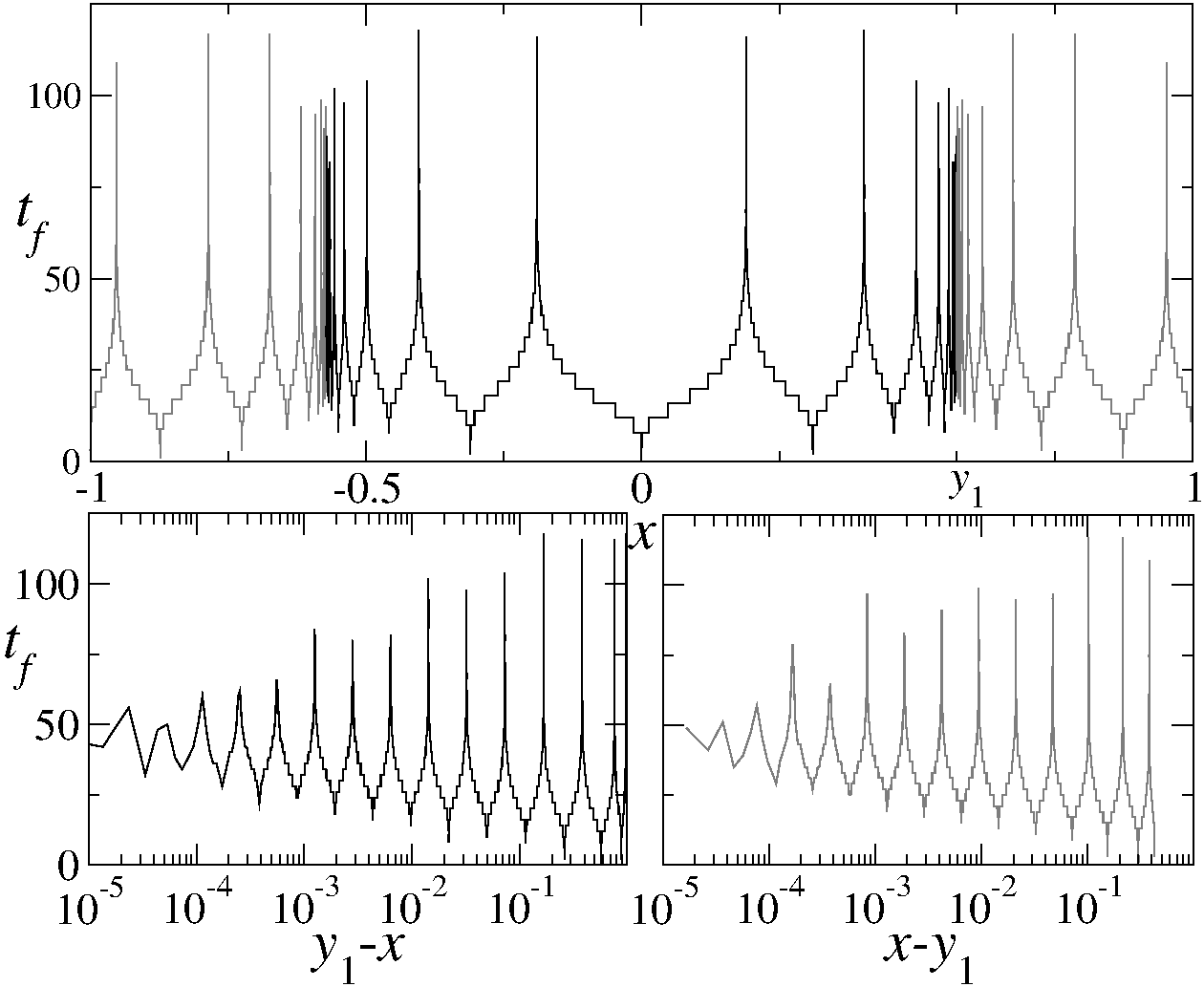}
\caption{\small (\textbf{Top panel}) Time of flight $t_{f}(x)$ for $N = 2$; the black lines correspond to initial conditions that terminate at the
attractor positions $x = 0$ and $x \simeq -0.310 703$ and the gray lines to trajectories ending at $x = 1$ and $x \simeq 0.8734$. (\textbf{Right (left) bottom panel}) Same as the top panel, but plotted against the logarithm of $x - y_{1} (y_{1} - x)$. It is evident that the peaks are arranged exponentially around the old repellor position, $y_{1}$, {\it i.e.}, they appear equidistant on a logarithmic scale.}
\label{fig:f4pre08}
\end{figure}

\begin{figure} 
\centering
\includegraphics[width=0.45\textwidth]{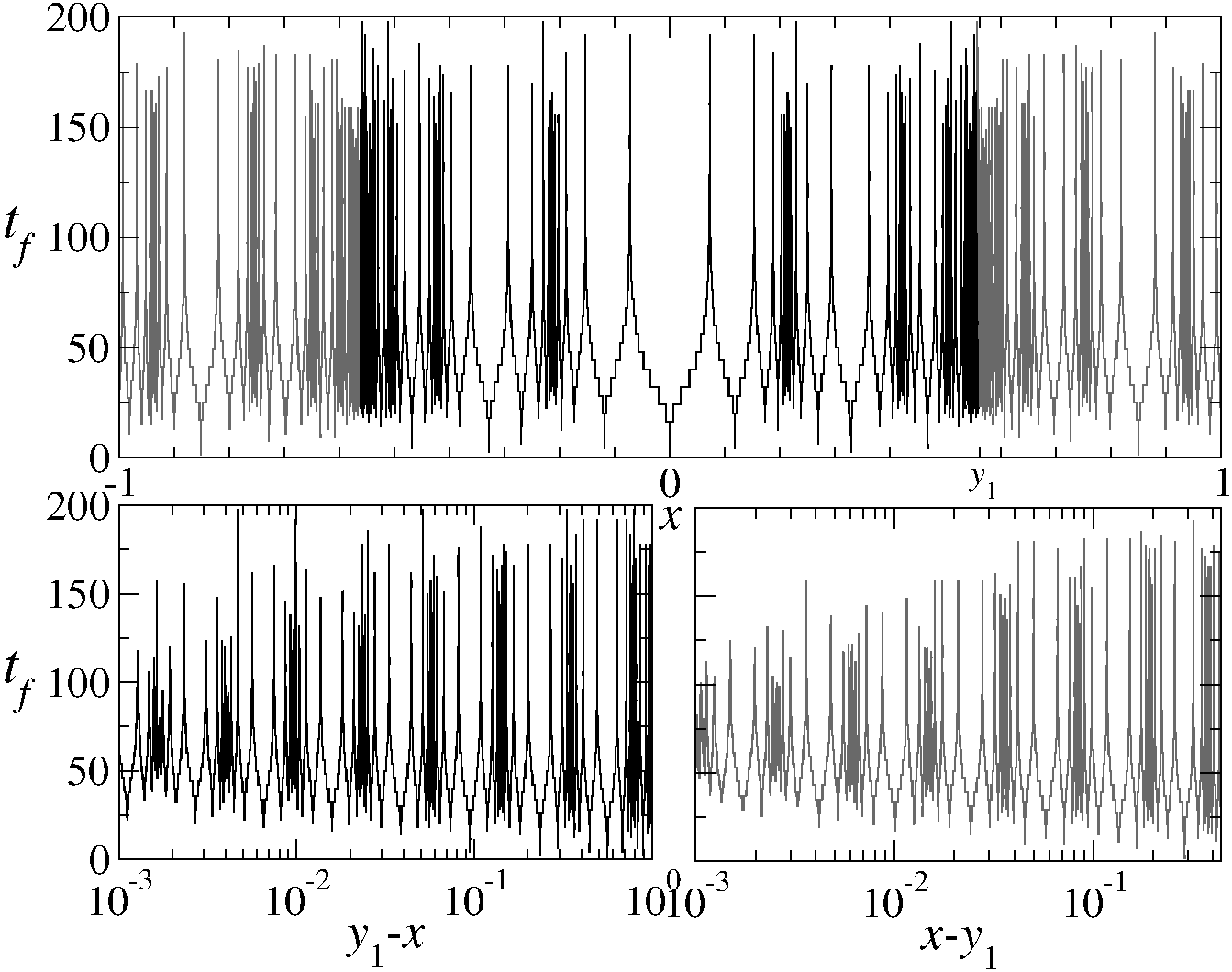}
\caption{\small Same as Figure \ref{fig:f4pre08}, but for $N = 3$. The black lines correspond to initial conditions that terminate at any of the four attractor positions close or equal to $x = 0$ and the gray lines to trajectories ending at any of the other four attractor positions close or equal to $x = 1$. As the bottom panels show, on a logarithmic scale, in this case, there are (infinitely) many clusters of peaks (repellor preimages) equidistant from each other}
\label{fig:f7pre08}
\end{figure}

The organization of the total set of trajectories as generated by all
possible initial conditions as they flow towards a period $2^{N}$ attractor is described in \cite{robledo2, robledo10}. It was
found that the paths taken by the full set of trajectories on their way to
the supercycle attractors (or to their complementary repellors) are
exceptionally structured. We define the preimage, $x^{(k)}$, of order $k$ of
position $x$ to satisfy $x=h^{(k)}(x^{(k)})$, where $h^{(k)}(x)$ is the $k$%
-th composition of the map, $h(x)\equiv f_{\overline{\mu }%
_{N}}^{(2^{N-1})}(x) $. The preimages of the attractor of period $2^{N}$, $%
N=1,2,3,...$, are distributed into different basins of attraction, one for
each of the $2^{N}$ phases (positions) that composes the cycle. When $N\geq 2$,
these basins are separated by fractal boundaries, whose complexity increases
with increasing $N$. The boundaries consist of the preimages of the
corresponding repellor and their positions cluster around the $2^{N}-1$
repellor positions, according to an exponential law. As $N$ increases, the
structure of the basin boundaries becomes more involved. Namely, the
boundaries for the $2^{N}$ cycle develops new features around those of the
previous $2^{N-1}$ cycle boundaries, with the outcome that a hierarchical
structure arises, leading to embedded clusters of clusters of boundary
positions, and so forth. The dynamics associated with families of trajectories
always displays a characteristically concerted order in which positions are
visited, which, in turn, reflects the repellor preimage boundary structure of
the basins of attraction. That is, each trajectory has an initial position
that is identified as a preimage of a given order of an attractor (or
repellor) position, and this trajectory necessarily follows the steps of
other trajectories, with the initial conditions of lower preimage order belonging
to a given chain or pathway to the attractor (or repellor). When the period $%
2^{N}$ of the cycle increases, the dynamics becomes more involved with
increasingly more complex stages that reflect the hierarchical structure of
preimages. See Figures \ref{fig:f4pre08} and \ref{fig:f7pre08}, and see \cite {robledo2, robledo10} for details.\ The fractal features of the
boundaries between the basins of attraction of the positions of the periodic
orbits develop a structure with hierarchy, and this, in turn, echoes in the
properties of the trajectories. The set of trajectories produce an ordered
flow towards the attractor or towards the repellor that follows through the ladder
structure of the sub-basins that constitute the mentioned boundaries.

From our previous discussion, we know that, every time the period of a
supercycle increases from $2^{N-1}$ to $2^{N}$ by a shift in the control
parameter value from $\overline{\mu }_{N-1}$ to $\overline{\mu }_{N}$, the
preimage structure advances one stage of complication in its hierarchy.
Along with this, and in relation to the time evolution of the ensemble of
trajectories, an additional set of $2^{N}$ smaller phase-space gaps develops,
and, also, a further oscillation takes place in the corresponding rate, $W_{t}$,
for finite period attractors. At $\mu _{\infty }$, the time evolution tracks
the period-doubling cascade progression, and every time $t$ increases from $%
2^{N-1}$ to $2^{N}$, the flow of trajectories undergoes equivalent passages
across stages in the itinerary through the preimage ladder structure, in the
development of phase-space gaps and in logarithmic oscillations in $W_{t}$.
Thus, each doubling of the period introduces additional modules or building
blocks in the hierarchy of the preimage structure, such that the complexity
of these added modules is similar to that of the total period $2^{N}$
system. As a consequence of this, we have obtained detailed understanding
of the mechanism by means of which the discrete scale invariance implied by
the log-periodic property \cite{robledo2} in the rate of approach $W_{t}$
arises.

\subsubsection{Dynamical Hierarchy}

We proceed now to the identification of the features of the dynamics towards
the Feigenbaum attractor as those of a bona fide model of dynamical
hierarchy with modular organization. These are: (i) elementary degrees of
freedom and the elementary events associated with them; (ii) building blocks
and the dynamics that takes place within them and through adjacent levels of
blocks; (iii) self-similarity characterized by coarse graining and
renormalization-group (RG) operations; and (iv) the emerging property of the entire
hierarchy absent in the embedded building blocks.

The elementary degrees of freedom are the preimages of the attractors of
period $2^{N}$. These are assigned an order $k$, according to the number $k$
of map iterations they require to reach the attractor. The preimages are
also distinguished in relation to the position, $x_{1}$, \ldots , $x_{2^{N}}$%
, of the attractor they reach first. The preimages of each attractor
position appear grouped in basins with fractal boundaries. The elementary
dynamical event is the reduction of the order $k$ of a preimage by one unit.
This event is generated by a single iteration of the map for an initial
position placed in a given basin. The result is the translation of the
position to a neighboring basin of the same attractor position.

The building blocks are clusters of clusters formed by families of boundary
basins. The (boundary) basins of attraction of the $2^{N}$ positions of the
periodic attractors cluster exponentially and have an alternating structure 
\cite{robledo2, robledo10}. In turn, these clusters cluster exponentially themselves
with their own alternating structure. Furthermore, there are clusters of
these clusters of clusters with similar arrangements, and so on. The
dynamics associated with the building blocks consists of the flow of
preimages through and out of a cluster or block. Sets of preimages ``evenly"
distributed (say, one per boundary basin) across a cluster of order $N$
(generated by an attractor of period $2^{N}$) flow orderly throughout the
structure. If there is one preimage in each boundary basin of the cluster,
each map iteration produces a migration from boundary basin to boundary
basin in such a way that, at all times, there is one preimage per boundary
basin, except for the inner ones in the cluster that are gradually emptied.

The self-similar feature in the hierarchy is demonstrated by the
coarse-graining property amongst building blocks of the hierarchy that
leaves the hierarchy invariant when $\mu =\mu _{\infty }$. That is, clusters
of order $N$ can be simplified into clusters of order $N-1$. There is a
self-similar structure for the clusters of any order, and a coarse graining
can be performed on clusters of order $N$, such that these can be reduced to
clusters of a lower order; the basic coarse graining is to transform order $N$
into order $N-1$. Furthermore, coarse graining can be carried out effectively via
the RG functional composition and rescaling, as this transformation reduces
the order $N$ of the periodic attractor. Automatically, the building-block
structure simplifies into that of the next lower order, and the clusters of
boundary basins of preimages are reduced in one unit of involvedness.
Dynamically, the coarse-graining property appears as flow within a cluster
of order $N$ simplified into flow within a cluster of order $N-1$. As coarse
graining is performed in a given cluster structure of order $N$, the flow of
trajectories through it is correspondingly coarse grained. e.g., flow out of
a cluster of clusters is simplified as flow out of a single cluster. RG
transformation via functional composition and rescaling of the cluster flow
is displayed dynamically, since by definition functional composition
establishes the dynamics of iterates, and the RG transformation $%
Rf(x)=-\alpha f(f(-x/\alpha )$ leads, for $N$ finite, to a trivial fixed point
that represents the simplest dynamical behavior, that of the period one
attractor. For $N\rightarrow \infty$, the RG transformation leads to the
self-similar dynamics of the non-trivial fixed point, the period-doubling
accumulation point.

There is a modular structure of embedded clusters of all orders. The
building blocks, clusters of order~$N$, form well-defined sets that are
embedded into larger building blocks, sets of clusters of order $N+1$. As
the period $2^{N}$ of the attractor that generates this structure increases,
the hierarchy extends, and as $N$ diverges, a fully self-similar structure
develops. The RG transformation for $N\rightarrow \infty$ no longer reduces
the order of the clusters, and a nontrivial fixed point arises. The
trajectories consist of embedded flows within clusters of all orders. The
entire flow towards the $2^{N}$-period attractor generated by an ensemble of
initial conditions (distributed uniformly across the phase space interval of
the map) methodically follows a pattern predetermined by the hierarchical
structure of embedded clusters of the~preimages.

Each module exhibits a basic kind of flow property. This is the exponential
emptying of trajectories within a cluster of order $N$. Trajectories
initiated in the boundary basins that form a cluster of order $N$ flow out
of it with an iteration time exponential law. The flow is transferred into a
cluster of order $N+1$. This flow is a dynamical module from which a
structure of flows is composed. The emerging property that appears when $%
N\rightarrow \infty$ is that there is a power-law emptying of trajectories
for the entire hierarchy. The flow of trajectories towards an attractor of
period $2^{N}$ proceeds via a sequence of stage or step flows, each within a
cluster of a given order. Thus, the first is through a cluster of order one,
then through a cluster of order two, \textit{etc}., until the last stage is through a
cluster of order $N$. The sequence evolves in time via a power law decay
that is modulated by logarithmic oscillations. This is the emerging property
of the model.

\section{Distributions of Sums of Deterministic Variables}

The limit distributions of sums of deterministic variables at the
period-doubling transition to chaos in unimodal maps have been studied by
making use of the trajectory properties described above \cite{fuentes1, fuentes1b}.
Firstly, the sum of positions as they are visited by a single internal
trajectory was found to have a multifractal structure imprinted by that of
the critical attractor \cite{fuentes1, fuentes1b}. It was shown, analytically and
numerically, that the sum of values of positions display discrete scale
invariance fixed jointly by Feigenbaum's universal constant, $\alpha$, and by
the period doublings contained in the number of summands. The stationary
distribution associated with this sum has a multifractal support given by
the Feigenbaum attractor. Secondly, the sum of subsequent positions
generated by an ensemble of uniformly distributed initial conditions in the
entire phase space was determined \cite{fuentes2}. It was found that this
sum acquires features of the repellor preimage structure that dominates the
dynamics toward the attractor. The stationary distribution associated with
this ensemble has a hierarchical structure with multifractal and discrete
scale invariance properties.

\subsection{Sums of Positions of a Single Trajectory within the Attractor}

\begin{figure} 
\centering
\includegraphics[width=0.5\textwidth]{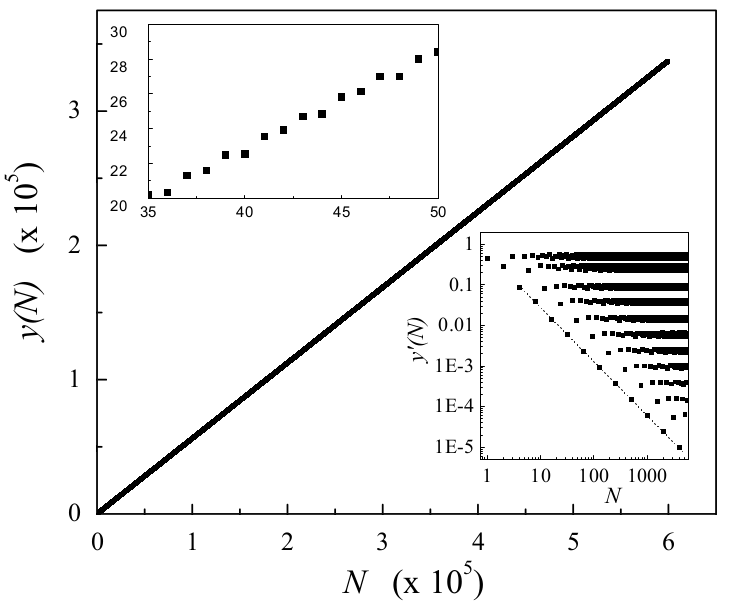}
\caption {
Sum of absolute values of visited points, $x_t, t =
0, ..., N $, of the Feigenbaum’s attractor with initial condition
$x_0 = 0$. Top inset: a closer look of the path of the sum,
for values of $N$ ranging between 35 and 50. Bottom inset:
Centered sum $y'(N)$ in logarithmic scales. See the text.}
\label{fig:onesum6}
\end{figure}

The starting point is the numerical evaluation of the sum of absolute values, 
$\left\vert x_{t}\right\vert$, at $\mu _{\infty}$ and with $x_{0}=0$:
\begin{equation}
y(N)\equiv \sum\limits_{t=1}^{N}\left\vert x_{t}\right\vert
\label{sumabs1}
\end{equation}%
Figure \ref{fig:onesum6} shows the result, where it can be observed that the values recorded, besides a repeating fluctuating pattern within a narrow band, increase linearly on the whole. The measured slope of the linear growth is $c=0.56245...$. The top inset shows an enlargement of the band, where some detail of the complex pattern of values of $y(N)$ is observed. A stationary view of the mentioned pattern is shown in bottom inset, where we plot \begin{equation}
y^{\prime }(N)\equiv \sum\limits_{t=1}^{N}\left( \left\vert x_{t}\right\vert
-c\right)
\label{sumabs2}
\end{equation}%
in logarithmic scales. There, we observe that the values of $y^{\prime }(N)$
fall within horizontal bands interspersed by gaps, revealing a fractal or
multifractal set layout. The top (zeroth) band contains $y^{\prime }$ for
all the odd values of $N$; the first band next to the top band contains $%
y^{\prime }$ for the even values of $N$ of the form $N=2+4m$, $m=0,1,2,...$.
The second band next to the top band contains $y^{\prime }(N)$ for $%
N=2^{2}+2^{3}m$, $m=0,1,2,...$, and so on. In general, the $k$-th band next
to the top band contains $y^{\prime }(2^{k}+2^{k+1}m)$, $m=0,1,2,...$.
Another important feature in this figure inset is that the $y^{\prime }(N)$
for subsequences of $N$ each of the form $N=(2l+1)2^{k}$, $k=0,1,2,...$,
with $l$ fixed at a given value of $l=0,1,2,...$, appear aligned with a
uniform slope $s=-1.323...$ (See the dotted line linking the values of $%
y^{\prime }(N)$ when $l=0$). The parallel lines formed by these subsequences
imply the power law $y^{\prime }(N)\sim N^{s}$ for $N$ belonging to such a
subsequence.

We have seen \cite{robledo1} that these two characteristics of $y^{\prime
}(N)$ are also present in the layout of the absolute value of the individual
positions $\left\vert x_{t}\right\vert $, $t=1,2,3,...$ of the trajectory
initiated at $x_{0}=0$; and this layout corresponds to the multifractal
geometric configuration of the points of the Feigenbaum's attractor; see Figure \ref{Fig._6}. In this case, the horizontal bands of positions separated by
equally-sized gaps are related to the period-doubling diameters, $d_{n,m}$,
that are used to construct the multifractal attractor. The identical slope
shown in the logarithmic scales by all the position subsequences $\left\vert
x_{t}\right\vert $, $t=(2l+1)2^{k}$, $k=0,1,2,...$, each formed by a fixed
value of $l=0,1,2,...$, implies the power law $\left\vert x_{t}\right\vert
\sim t^{s}$, $s=-\ln \alpha /\ln 2=-1.3236...$, as the $\left\vert
x_{t}\right\vert$ can be expressed as:%
\begin{equation}
\left\vert x_{t}\right\vert \simeq \left\vert x_{2l+1}\right\vert \ \alpha
^{-k}
\label{powerlaw1}
\end{equation}%
$t=(2l+1)2^{k}$, $k=0,1,2,...$, or, equivalently, $\left\vert
x_{t}\right\vert \sim t^{s}$. Notice that the index $k$ also labels the
order of the bands from top to bottom. The power law behavior involving the
universal constant, $\alpha$, of the subsequence positions reflect the
approach of points in the attractor toward its most sparse region at $x=0$
from its most compact region, as the positions at odd times $\left\vert
x_{2l+1}\right\vert =x_{2l+1}$, those in the top band, correspond to the
densest region of the set. Having uncovered the manifestation of the
multifractal structure of the attractor into the sum, $y^{\prime }(N)$, it is
possible to proceed to derive Equations (\ref{sumabs1}) and (\ref{sumabs2})
analytically \cite{fuentes1, fuentes1b} and to corroborate that the value of the slope, $s$, in Figure \ref{fig:onesum6} is indeed given by $s=-\ln \alpha /\ln 2=-1.3236...$. The
derivation also illustrates the discrete scale invariant property of the
time evolution at the period doubling transition to chaos. That is, a
duplication in iteration time is accompanied by a scale factor in the
iterate position equal to the universal constant, $\alpha$.

In \cite{fuentes1, fuentes1b}, the next step was to consider the straight sum of $%
x_{t}$, where the signs taken by positions lessen the growth of its value as 
$N$ increases, and the results found were consistently similar to those for
the sum of $\left\vert x_{t}\right\vert $, {\it i.e.}, the linear growth of a
fixed-width band within which the sum displays a fluctuating arrangement.
Numerical and analytical details for the sum of $x_{t}$ are described in
detail in \cite{fuentes1, fuentes1b}. After this, numerical results for the sum of
iterated positions obtained when the control parameter is shifted into the
region of chaotic bands were obtained. In all of these cases, the
distributions evolve after a characteristic crossover towards a Gaussian
form. These findings were explained in terms of an RG framework in which the
action of the central limit theorem plays a fundamental role and provide
details of the crossover from multiband distributions to the Gaussian
distribution \cite{fuentes1, fuentes1b}.

\subsection{Sums of Positions of an Ensemble of Trajectories Evolving towards the Attractor}

\begin{figure} 
\centering 
\includegraphics[width=0.5\textwidth]{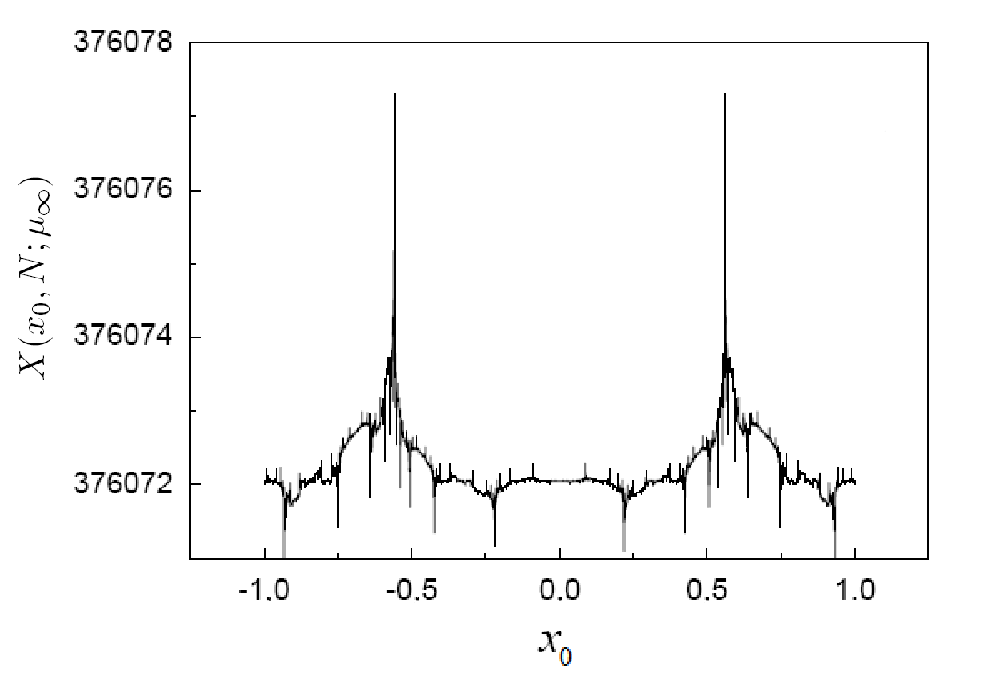}
\caption{\small Sums $X(x_{0},N;\mu_{\infty})$ as a function of $x_{0}$, $N\sim O(10^{6})$ at the Feigenbaum point.}
\label{sumsd}
\end{figure}

\begin{figure} 
\centering 
\includegraphics[width=0.5\textwidth]{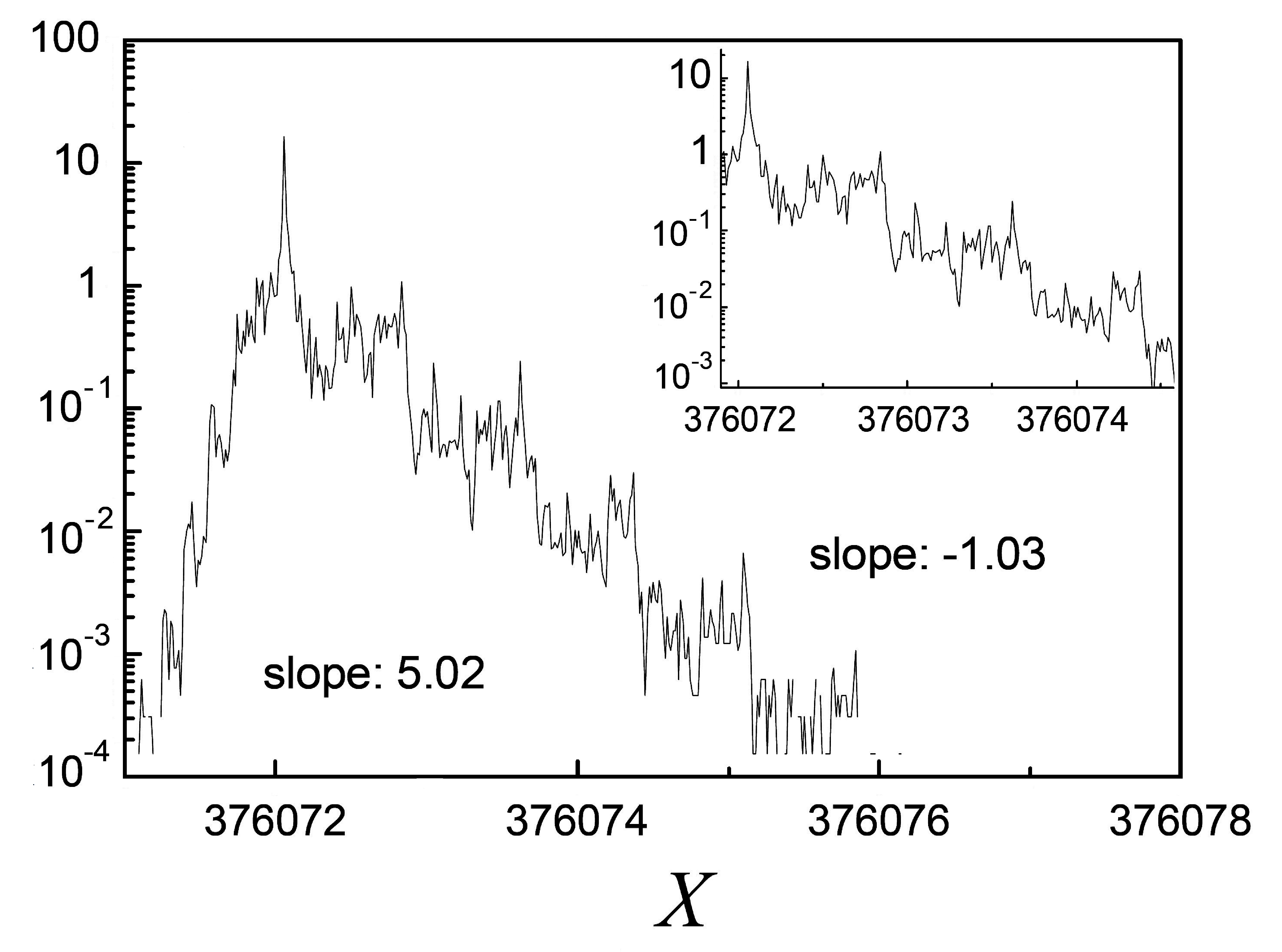}
\caption{\small Histograms obtained from the sums in Figure \ref{sumsd}. The inset shows greater detail.}
\label{distd}
\end{figure}

A second, different, but complementary, study \cite{fuentes2} consists of the evaluation of the sum of positions, $x_{t}$, up to a final iteration time, $N$, of a trajectory with initial condition $x_{0}$ and control parameter value $\mu $, {\it i.e.},:
\begin{equation}
X(x_{0},N;\mu )\equiv \sum\limits_{t=0}^{N}x_{t}
\end{equation} 
Figure \ref{sumsd} shows the results for $X(x_{0},N;\mu _{\infty}$ for all possible initial conditions, $
-1\leq x_{0}\leq 1$ and $N\sim O(10^{6})$. The plot is symmetrical with respect to $x_{0}=0$ and exhibits two large peaks and a central valley. This provides the main frame onto which motifs
of alternating signs and diminishing amplitude are added consecutively.
The distribution associated with the set of sums, $X(x_{0},N;\mu _{\infty}$, is shown in Figure~\ref{distd}, which we observe has an asymmetrical double exponential global shape with
superposed motifs of ever-decreasing finer detail characteristic of a
multifractal object. The dynamical properties described above of trajectories as
they evolve towards the Feigenbaum attractor can be used to explain the dynamical origin of all the features of this distribution.
Bearing in mind the basic property that trajectories at $\mu _{\infty }$
from $t=0$ up to $t\rightarrow \infty$ trace the period-doubling cascade
progression from $\mu=0$ up to $\mu_{\infty }$, it is clear now how to decode the structure of the sums in
Figure \ref{sumsd} and that of its histogram in Figure \ref{distd}. As explained in \cite{fuentes2}, the arrangement of
the multiscale families of cusps of $X(x_{0},N;\mu _{\infty })$ is the
manifestation of the consecutive formation of phase space gaps in an
initially uniform distribution of positions $x_{0}$ and the logarithmic
oscillations at times $t^{k}$, $k=1,2,3,...$ of the rate of convergence of
trajectories $W_{t}$ to the Feigenbaum attractor. The mean time for the
opening of gaps of the same order is $t^{k}$, $k=1,2,3,...$, and the signs
and amplitudes of the cusps are the testimonies stamped in the sums of the
main passages out of the labyrinths formed by the preimages of the repellor,
the transits of trajectories from one level of the hierarchy to the next.
Visual representation of the stationary distribution, $P(X,N\rightarrow
\infty ;\mu_{\infty })$, associated with the sums, $X(x_{0},N;\mu _{\infty })$, is shown
by the histogram in Figure \ref{distd}. It has an asymmetrical double exponential
backbone onto which multiscale patterns are attached that originate from the
aforementioned sets of cusps in the sums. The negative slope (in
semi-logarithmic scales) of the backbone originates from the main $k=1$
repellor core structure shown for all supercycles, while the steeper positive
slope originated form the replica structures of the former that originate
from the $k=2$ repellor structures.

We have found that the sum of iterate positions \textit{within} the 
Feigenbaum's attractor has a multifractal structure stamped by that of the
attractor, while that involving positions of an ensemble of trajectories
moving \textit{toward} the attractor exhibits the preimage structure of its
corresponding repellor. These basic features suggest a degree of
universality, limited to the critical attractor under consideration, in the
properties of sums of deterministic variables at the transitions to chaos.
Namely, the sums of positions of memory-retaining trajectories evolving
under a vanishing Lyapunov exponent appear to preserve the particular
features of the multifractal critical attractor and repellor under
examination. Thus, we expect that varying the degree of nonlinearity of a
unimodal map would affect the scaling properties of time averages of
trajectory positions at the period doubling transition to chaos, or
alternatively, that the consideration of a different route to chaos, such as
the quasiperiodic route, would lead to different scaling properties of
comparable time averages. Furthermore, as commented on above, the spiked functional
dependence of the sums $X(x_{0},N;\mu_{\infty})$ on $x_{0}$ shown 
in Figure \ref{sumsd} follows the characteristic hierarchical preimage
structure, with exponential clustering, around the positions of the major
and other high ranking elements of the repellor and their first preimages 
\cite{robledo2}. This feature suggests that the distribution of large sums
of positions is dominated by long journeys toward the attractor that are
particular to the critical attractor under consideration.

\section{Manifestations of Incipient Chaos in Condensed Matter Systems}

It is of interest to know if the anomalous dynamics found for critical
attractors in low-dimensional maps bears some correlation with the anomalous
dynamical behavior at extremal or transitional states in systems with many
degrees of freedom characteristic of condensed matter physics. Three
specific examples have been developed. In one case, the dynamics at the
tangent bifurcation has been shown to be related to that of evanescent
fluctuations, or order parameter clusters, at thermal critical states \cite%
{robcrit1, robcrit1b}. The second case also involves the tangent bifurcation and
appears in a description of the localization transition between a conducting
and an insulating phase \cite{robloc1}. In the third case, the dynamics at
the period-doubling onset of chaos has been demonstrated to be closely
analogous to the glassy dynamics observed in supercooled molecular liquids 
\cite{robglass1, robglass1b, robglass1c}.

\subsection{Critical Clusters}

We examine first the intermittent properties of clusters of the order parameter
at critical thermal states \cite{robcrit1, robcrit1b}. Our interest is in the local
fluctuations of a system undergoing a second order phase transition, for
example, in the Ising model as the magnetization fluctuates and generates
magnetic domains on all size scales at its critical point. In particular,
the object of study is a single cluster of order parameter $\phi$ at
criticality. This is described by a coarse-grained free energy or effective
action, like in the Landau--Ginzburg--Wilson (LGW) continuous spin model
portrayal of the equilibrium configurations of Ising spins at the critical
temperature and zero external field. At criticality, the LGW free energy
takes the form:
\begin{equation}
\Psi _{c}[\phi ]=a\int dr^{d}\left[ \frac{1}{2}\left( \nabla \phi \right)
^{2}+b\left\vert \phi \right\vert ^{\delta +1}\right]  \label{lgw1}
\end{equation}%
Where $a$ and $b$ are constants, $\delta$ is the critical isotherm exponent
and $d$ is the spatial dimension ($\delta =5$ for the $d=3$ Ising model with
short range interactions). As we describe below, a cluster of radius $R$ is
an unstable configuration whose amplitude in $\phi$ grows in time and
eventually collapses when an instability is reached. This process has been
shown \cite{athens1, athens2} to be reproduced by a nonlinear map
with tangency and feedback characteristics, such that the time evolution of
the cluster is given in the nonlinear system as a laminar event of
intermittent dynamics.

The method employed to determine the cluster's order parameter profile, $\phi
(\mathbf{r})$, adopts the saddle-point approximation of the coarse-grained
partition function, $Z$, so that $\phi (\mathbf{r})$ is its dominant
configuration and is determined by solving the corresponding Euler--Lagrange
equation. The procedure is equivalent to the density functional approach for
stationary states in equilibrium nonuniform fluids. Interesting properties
have been derived from the thermal average of the solution found for $\phi (%
\mathbf{r})$, evaluated by integrating over its amplitude, $\phi _{0}$, the
remaining degree of freedom after its size, $R$, has been fixed. These are the
fractal dimension of the cluster \cite{athens3, athens4} and the
intermittent behavior in its time evolution \cite{athens1, athens2}.
Both types of properties are given in terms of the critical isotherm
exponent, $\delta $.

As we describe below, the dominance of $\phi (\mathbf{r})$ in $Z$ depends on
a condition that can be expressed as an inequality between two lengths in
space. This is $r_{0}\gg R$, where $r_{0}$ is the location of a divergence
in the expression for $\phi (\mathbf{r})$ that decreases as an inverse power
of the cluster amplitude, $\phi _{0}$. When $r_{0}\gg R$, the profile is
almost horizontal, but for $r_{0}\gtrapprox R$, the profile increases from its
center faster than an exponential. This is responsible for the cluster properties described in \cite{robcrit1, robcrit1b}. These properties relate to
the dependence of the number of cluster configurations on size $R$ and the
sensitivity to initial conditions $\xi _{t}$ of the order-parameter evolution on
time $t$. When the method considers a one dimensional system with
unspecified range of interactions, the magnetization profile turns out
exactly to have the form of a $q$-exponential:
\begin{equation}
\phi (x)=\phi _{0}\exp _{q}(kx)
\end{equation}
with $q=$ $(1+\delta )/2$ and $k=\sqrt{2b}\phi _{0}^{(\delta -1)/2}$ \cite%
{robcrit1, robcrit1b}. Because $\delta >1$, one has $q>1$, and $\phi (x)$ grows faster
than an exponential as $x\rightarrow x_{0}$ and diverges at $x_{0}$. It is
important to notice that only configurations with $R\ll x_{0}$ have a
nonvanishing contribution to the path integration in $Z$ and that these
configurations vanish for the infinite cluster-sized system. There are some
characteristics of nonuniform convergence in relation to the limits, $%
R\rightarrow \infty$ and $x_{0}\rightarrow \infty $, a feature that is
significant for our connection with $q$-statistics. By taking $\delta =1$,
the system is set out of criticality; then, $q=1$, and the profile, $\phi (x)$,
becomes the exponential $\phi (x)=\phi _{0}\exp (k_{0}\ x)$, $k_{0}=\sqrt{%
a_{0}t}$. The analysis can be carried out for higher dimensions with no
further significant assumptions and with comparable results \cite{athens1, athens2, athens3, athens4}.

\begin{figure} 
\centering
\includegraphics[width=0.5\textwidth]{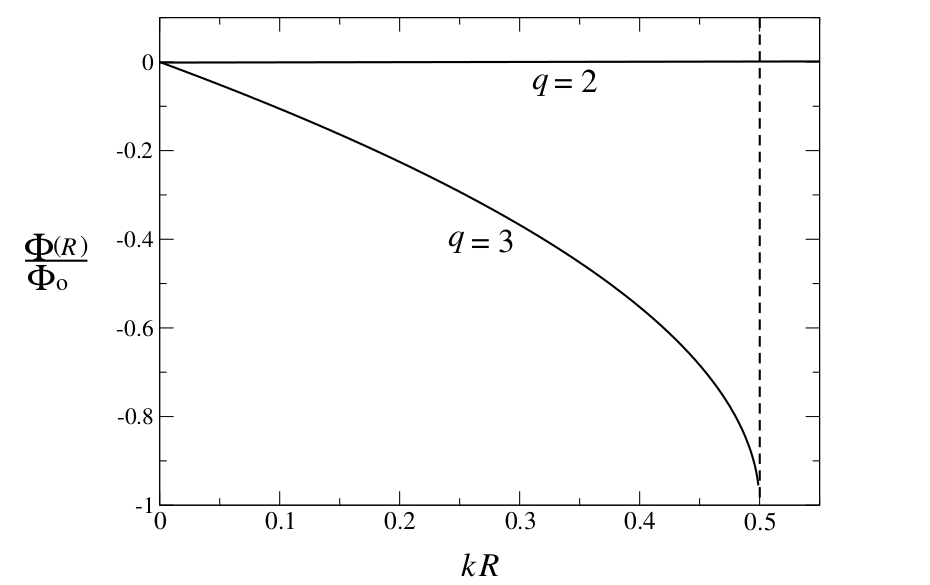}
\caption{Total magnetization of the cluster, $\Phi (R)$, as a function of the cluster size, $R$, according to Equation (\ref{eq:totmag}) for $q=2$ and $3$, which correspond, respectively, to $\delta=3$ and $5$.}
\label{fig:magtot}
\end{figure}

The total magnetization of the cluster of size $R$ (see Figure \ref{fig:magtot}):

\begin{equation}
\Phi (R)=\int_{0}^{R}dx\phi (x) = \Phi _0 \lbrace [\exp _q (kR)]^{2-q}-1 \rbrace ,\ R < x_0
\label{eq:totmag}
\end{equation}%
can be used to estimate its entropy rate of growth with $R$ \cite{robcrit1, robcrit1b},
and it is significant to note that this rate has the form of the Tsallis
entropy and complies with the extensivity property, $S_{q}\sim R$, while the
BG entropy obtained from $S_{q}$ when $q=1$ complies also with the
extensivity property, $S_{1}\sim R$~\cite{robcrit1, robcrit1b}. That is, $S_{q}$ is
extensive only at criticality, and the BG $S_{1}$ is extensive only off
criticality. A condition for these properties associated with $S_{q}$, $q>1$,
to arise is criticality, but also is the situation that phase space has only
been partially represented by selecting configurations that are dominant only under certain conditions. Hence,
the motivation to examine this problem rests on explaining the physical and
methodological basis under which proposed generalizations of the BG
statistics may apply.

\subsection{Mobility Edge}

\begin{figure} 
\centering
\includegraphics[width=0.5\textwidth]{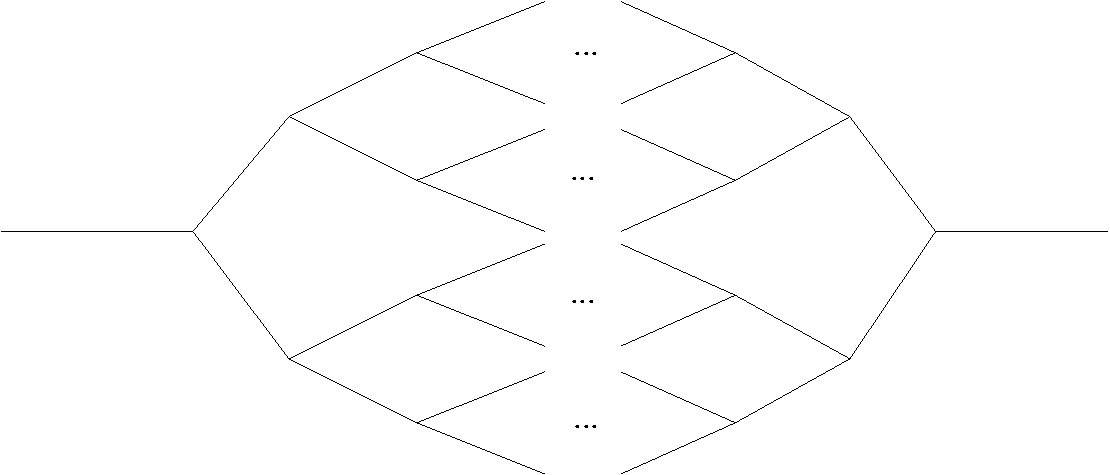}
\caption{A double Cayley tree of connectivity $K = 2$ and lattice
constant $a$. Each bond is a perfect one-dimensional conductor.}
\label{fig:cayleyd1}
\end{figure}

We refer now to a remarkable equivalence between the dynamics of an
intermittent nonlinear map and the electronic transport properties (obtained
via the scattering matrix approach) of a crystal defined on a double Cayley tree \cite
{robloc1}. See Figure \ref{fig:cayleyd1}. This analogy is strict \cite{robloc1} and reveals in detail the
nature of the mobility edge normally studied near, but not at, the
metal-insulator transition in disordered systems. An analytical expression
for the conductance that has a $q$-exponential form was obtained as a function of system size at this
transition. This manifests as power-law decay
or fast variability according to different kinds of boundary
conditions. The model does not
contain disorder; nevertheless, it displays a transition between localized
and extended states. The translation of the map dynamical expressions into
electronic transport terms provides not only the description of two
different conducting phases, but offered a
rigorous account of the conductance at the mobility edge \cite{robloc1}. A
new type of localization length in the incipient insulator mirrors the
departure from exponential sensitivity to initial conditions at the
transition to chaos.

\begin{figure} 
\centering
\includegraphics[width=0.5\textwidth]{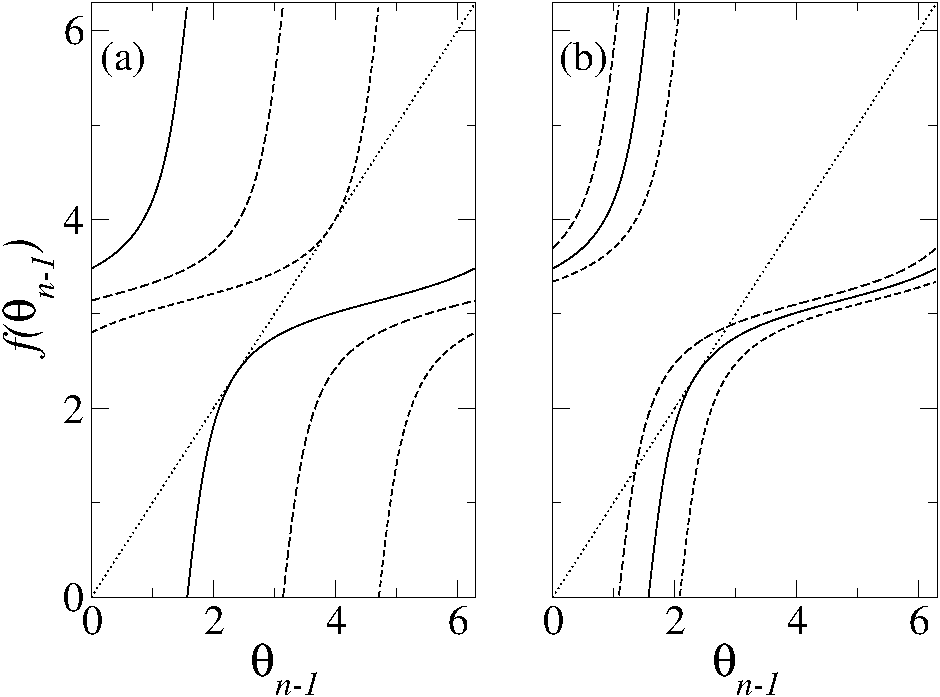}
\caption{\small $f(\theta_{n})$ of Equation (\ref{eq:map2a}), for $\epsilon = \frac{1}{4}$: (\textbf{a}) $ka= \pi /4$ (dashed), $2\pi/4$ (long dashed), $3\pi/4$ (continuous); and (\textbf{b}) $ka=2.6$ (long dashed), $3\pi/4$ (continuous), 2.1 (dashed). The dotted lines correspond to the identity.} 
\label{fig:maploc}
\end{figure}

The recursion formula for scattering matrices corresponding to consecutive sizes was reduced to a
nonlinear map, where time iterations represent increments in size. The number
of times the trees are ramified, starting from the perfect wire or lead, is the
generation, $n$, that quantifies the size of the system. For brevity, the tree connectivity was fixed to $K=2$, where one lead, the incoming lead,
is divided into two outgoing leads at a given node. Basic to the
discussion is the fact that the scattering matrix recursive relation can be
written in a diagonal form, and this implies the existence of a
one-dimensional nonlinear map for the electronic phase, $\theta _{n}$. The
map $\theta _{n+1}=f(\theta _{n})$ was obtained in the following
closed form \cite{robloc1}:
\begin{equation}
f(\theta _{n})=2ka-\theta _{n}+2\arctan \left( \frac{\sin \theta _{n}+\sqrt{%
1-2\epsilon }\sin 2ka}{\cos \theta _{n}-\sqrt{1-2\epsilon }\cos 2ka}\right) 
\label{eq:map2a}
\end{equation}
where the dependence on the transmission probability, $\epsilon $, the $k$
momentum and the lattice constant, $a$, comes out explicitly. Figure \ref{fig:maploc} illustrates the map in Eq. (\ref{eq:map2a}) for different parameter values.

\begin{figure} 
\centering
\includegraphics[width=0.5\textwidth]{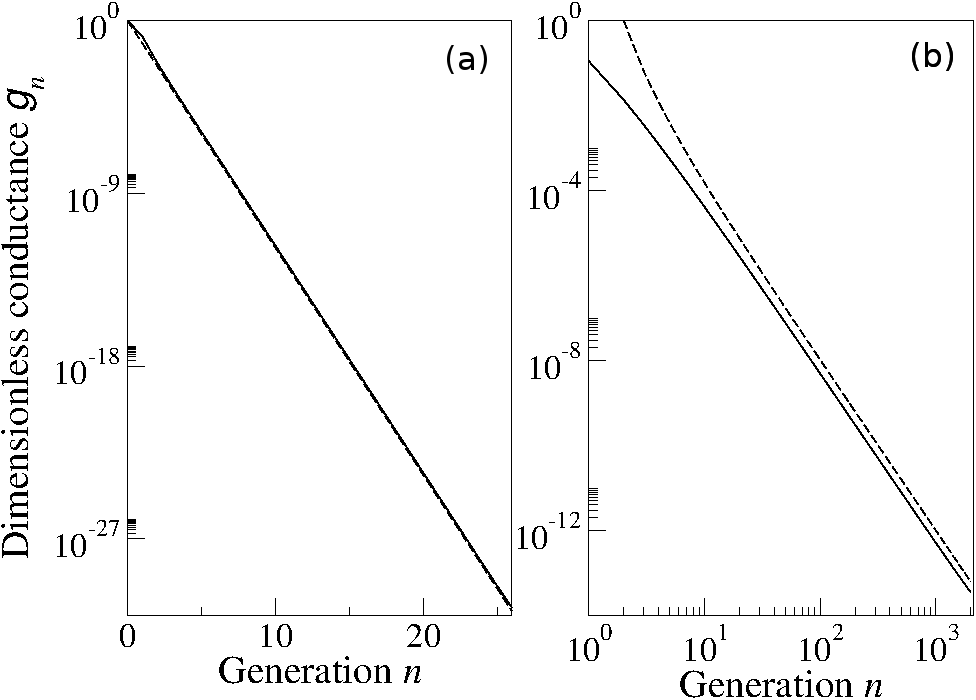}
\caption{\small Conductance as a function of generation for $\protect\epsilon =%
\frac{1}{4}$, (\textbf{a}) $ka=0.5$ and (\textbf{b}) $ka=\protect\pi /4$. Continuous lines
represent $g_{n}$ obtained directly from the map through Equation (\protect\ref%
{eq:map2a}), while dotted lines correspond to $g_{n}=\exp (2\protect\lambda %
_{1}n)$ and $g_{n}=\exp _{3/2}(2\protect\lambda _{3/2}n)$ for panels (a) and
(b), respectively. The two curves in (b) differ because of a proportionality
factor; see Equation (\ref{eq:transition}).}
\label{fig:gn}
\end{figure}

The dynamical properties of the map given by Equation (\ref{eq:map2a}) translate into the
following network properties: (i) There are two families of period one attractors for small and large values of $ka$ separated by a family of chaotic attractors for intermediate values of $ka$. For fixed $\epsilon$, these families are connected by tangent bifurcation points at $k_c a$ and $k'_c a$. (ii) In relation to the attractors of period one
that take place along $0\leq ka<k_{c}a$ and $k' _{c}a < ka\leq \pi$, it was found that the
ordinary Lyapunov exponent, $\lambda _{1}$, is negative, and therefore, the
conductance, $g_{n}$, decays exponentially with system size $n$, $g_{n}=\exp
(2\lambda _{1}n)$, implying localization, the localization length being $%
\zeta _{1}=a/|\lambda _{1}|$. In Figure \ref{fig:gn}a, we see a clear
exponential decay of $g_{n}$ as a function of $n$ at $ka=0.5$, where we
compare $g_{n}$ computed directly with that obtained from $\lambda _{1}$. (iii) With respect to the chaotic
attractors that occur in the interval, $k_{c}a<ka<k' _{c}a$, it was
observed that $\lambda _{1}$ becomes positive, and the recursion relation
does not let $g_{n}$ decay, but makes it oscillate with $n$ (not shown here),
indicating that conduction takes place. In our model, $g_{n}$ does not scale
with system size as in the metallic regime of quasi-one-dimensional
disordered wire, where Ohm's law is satisfied. In the parameter region where
the map is incipiently chaotic, say $ka\gtrsim k_{c}a$, the network grows
with $n$ with an insulator character, but interrupted for other intermediate
values of $n$ by conducting crystals. In the map dynamics, these are the
laminar episodes separated by chaotic bursts in intermittent trajectories.

The most distinct outcome of the treatment in \cite{robloc1} is the
description obtained of the mobility edge from the dynamics at the critical
points located at $ka=k' _{c}a$ and $ka=k_{c}a$. There, $\zeta_1 = a/|\lambda _1| \rightarrow \infty$, and not much can be said about
the size dependence of the conductance when $n\gg 1$. However, we can use to
our advantage the known properties of the anomalous dynamics occurring at
these transitions once they are identified as tangent bifurcations (see Figure \ref{fig:gn}b for $ka=3\pi /4$). As we have seen, at a tangent bifurcation of general
nonlinearity $z>1$, the sensitivity obeys a $q$-exponential law for large $n$~\cite{baldovin1}:
\begin{equation}
\xi _{n}=\exp _{q}(\lambda _{q}n)\equiv \lbrack 1-(q-1)\lambda _{q}n]^{\mp 
\frac{1}{q-1}}
\label{eq:q-sensitivity}
\end{equation}
where $\lambda _{q}$ is a $q$-generalized Lyapunov coefficient given by $%
\lambda _{q}=\mp zu$, $q=2-1/z$, where $u$ is the leading term of the
expansion up to order $z$ of $\theta _{n}$ close to $\theta _{c}$ (or $%
\theta ' _{c}$); {\it i.e.}, $\theta _{n}-\theta _{c}=(\theta _{n-1}-\theta
_{c})+u\,|\theta _{n-1}-\theta _{c}|^{z}+\cdots $. The minus and plus signs
in Equation (\ref{eq:q-sensitivity}) and in $\lambda _{q}$ correspond to
trajectories at the left and right, respectively, of the point of tangency, $%
\theta _{c}$. Equation (\ref{eq:q-sensitivity}) implies the power-law decay of $\xi
_{n}$ with $n$ when $\theta _{n}-\theta _{c}<0$ and faster than exponential
growth when $\theta _{n}-\theta _{c}>0$. By making the expansion around $%
\theta _{c}$ (or $\theta ' _{c}$) for map Equation (\ref{eq:map2a}), it
was found (as anticipated) $z=2$, implying $q=3/2$, and $u=\sqrt{%
(1-2\epsilon )/2\epsilon }$. The conductance $g_{n}$ at each bifurcation point acquires the form $g_{n}=\exp _{3/2}(2\lambda _{3/2}n)$ \cite{robloc1}, or:
\begin{equation}
g_{n}\propto \left( 1-\frac{1}{2}\lambda _{3/2}n\right) ^{-4}
\label{eq:transition}
\end{equation}%
where the $q$-generalized Lyapunov exponent is $\lambda _{3/2}=-2\sqrt{%
1-2\epsilon /2\epsilon }<0$. In the right panel of Figure \ref{fig:gn}b, results are compared for the conductance obtained from Equation (\ref{eq:transition}) (dashed line) and that obtained from the Landauer formula \cite{robloc1}. 
It is clear that $g_{n}$ decays as a power law (with quartic
exponent) rather than the exponential in the insulating phase. We emphasize
that a localization length given by $\zeta _{3/2}=a/\lambda _{3/2}$ can
still be defined at the mobility edge. When $\theta _{n}-\theta _{c}>0$, $\lambda _{3/2}=2%
\sqrt{1-2\epsilon /2\epsilon }>0$ and the recursion relation for $g_{n}$
describes an incipient conductor with exceptionally fast fluctuating $g_{n}$.

In summary, we can draw significant conclusions about electronic transport
from the study in \cite{robloc1}. These arise naturally when
considering the dynamical properties of the equivalent nonlinear map near or
at the intermittency transition to chaos. Since iteration time in the map
translates into the generation, $n$, of the network, time evolution means
the growth of system size, reaching the thermodynamic limit (and true self-similarity) when $n\rightarrow \infty $. In that limit, windows of period
one separated by a chaotic band correspond, respectively, to localized and
extended electronic states. Further, in the referred parameter ($ka$, $%
\epsilon $) regions, the conductance, $g_{n}$, of the model crystal shows
either an exponential decay with system size, with localization length given
by $\zeta _{1} = 1/|\lambda _1|$ (as in the case of a quasi-one-dimensional disordered wire),
or a fluctuating property, signaling conducting states. The pair of tangent
bifurcation points of the map correspond to the band or mobility edges that
separate conductor from insulator behavior. At these bifurcations, the
sensitivity to initial conditions, $\xi _{n}$, exhibits either power-law decay
(when $\theta _{0}<\theta ' _{c}$ or $\theta _{0}>\theta _{c}$) or
faster than exponential increase ($\theta ' _{c}<\theta _{0}<\theta
_{c}$), and consequently, the conductance inherits comparable decay or
variability with system size $n$. Notably, at the mobility edge, it is still
possible to define a localization length, the $q$-generalized localization,
length $\zeta _{q} = 1/|\lambda _q|$, with a fixed value of $q=3/2$. This expression is
universal, {\it i.e.}, it is satisfied by all maps that in the neighborhood of the
point of tangency have a quadratic term, {\it i.e.}, $z=2$ \cite{baldovin1}.

\subsection{Glassy Dynamics}

A third example of a link between critical attractors and condensed-matter
systems properties is the realization \cite{robglass1, robglass1b, robglass1c} that the dynamics at
the noise-perturbed period-doubling onset of chaos in unimodal maps is
analogous to that observed in supercooled liquids close to vitrification.
Four major features of glassy dynamics in structural glass formers, two-step
relaxation, aging, a relationship between relaxation time and
configurational entropy, and evolution from diffusive behavior to arrest,
are shown to be displayed by the properties of orbits with a vanishing
Lyapunov coefficient. The previously known properties in control-parameter
space of the noise-induced bifurcation gap \cite{schuster1, crutchfield1} play a central role in determining the characteristics of
dynamical relaxation at the chaos threshold.

A very pronounced slowing down of relaxation processes is the principal
expression of the approach to the glass transition \cite{debenedetti1, debenedetti2}, and this is generally interpreted as a progressively more
imperfect realization of phase space sampling. Because of this extreme
condition, an important question is to find out whether, under conditions of
ergodicity and mixing breakdown, the BG\ statistical mechanics remains
capable of describing stationary states in the immediate vicinity of glass
formation.

The basic ingredient of ergodicity failure is obtained for orbits at the
onset of chaos (at $\mu _{\infty }$ for unimodal maps) in the limit towards
vanishing noise amplitude. Our study supports the idea of a degree of
universality underlying the phenomenon of vitrification and points out that
it is present in different classes of systems, including some with no
explicit consideration of their molecular structure. The map has only one
degree of freedom, but the consideration of external noise could be taken to
be the effect of many other systems coupled to it, like in the so-called
coupled map lattices \cite{kaneko1}. Our interest has been to study a system
that is gradually forced into a non-ergodic state by reducing its capacity to
sample regions of its phase space that are space filling, up to a point at
which it is only possible to move within a multifractal subset of this
space. The logistic map with additive external noise reads: 
\begin{equation}
x_{t+1}=f_{\mu }(x_{t})=1-\mu x_{t}^{2}+\sigma \chi _{t},\;-1\leq x_{t}\leq
1,0\leq \mu \leq 2
\label{logistic1}
\end{equation}%
where $\chi _{t}$ is a Gaussian-distributed random variable with average $%
\left\langle \chi _{t}\chi _{t^{\prime }}\right\rangle =\delta _{t.t^{\prime
}}$, and $\sigma$ measures the noise intensity \cite{schuster1, crutchfield1}.

The erratic motion of a Brownian particle is usually described by the
Langevin theory \cite{chaikin1}. As is well known, this method finds a
way to avoid the detailed consideration of many degrees of freedom by
representing with a noise source the effect of collisions with molecules in
the fluid in which the particle moves. The approach to thermal equilibrium
is produced by random forces, and these are sufficient to determine
dynamical correlations, diffusion and a basic form for the
fluctuation-dissipation theorem~\cite{chaikin1}. In the same spirit,
attractors of nonlinear low-dimensional maps under the effect of external
noise can be used to model states in systems with many degrees of freedom.
Notice that the general map formula:
\begin{equation}
x_{t+1}=x_{t}+h_{\mu }(x_{t})+\sigma \chi _{t}
\end{equation}%
is a discrete form for a Langevin equation with the nonlinear ``friction force"
term, $h_{\mu }$, and $\chi _{t}$ is the same Gaussian white noise random
variable as in Equation (\ref{logistic1}) and $\sigma$, the noise intensity. With
the choice $h_{\mu }(x)=1-x-\mu x^{2}$, we recover Equation (\ref{logistic1}).


\subsubsection{Noise-Perturbed Onset of Chaos}

\begin{figure} 
\begin{center}
\includegraphics[width=0.45\textwidth]{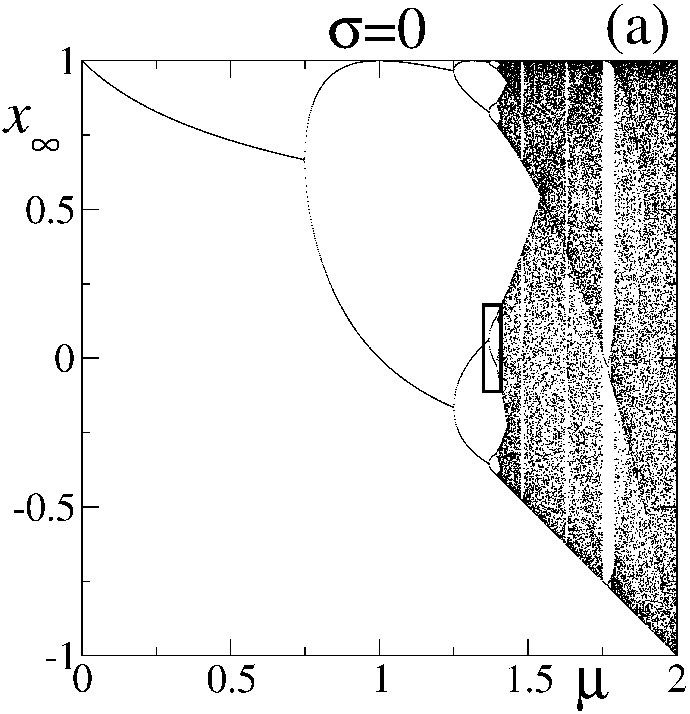} %
\includegraphics[width=0.45\textwidth]{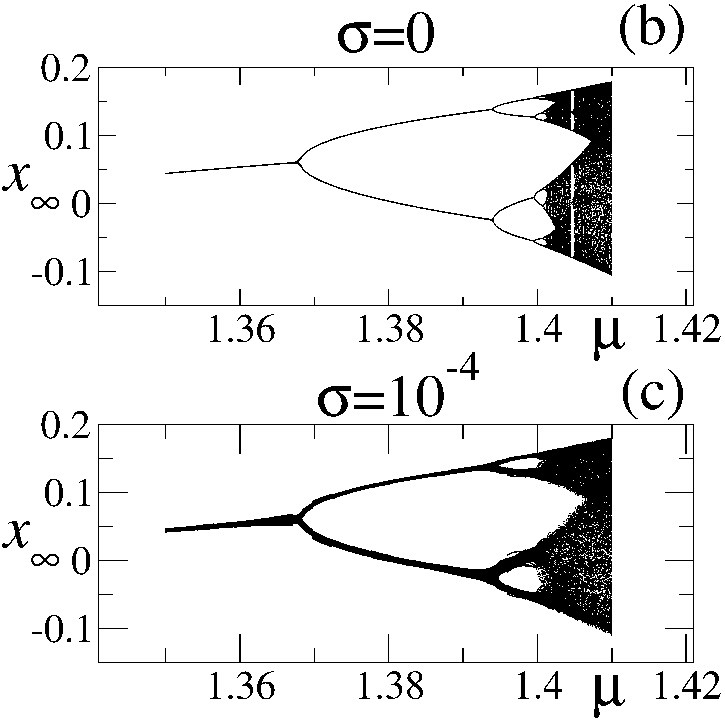}
\end{center}
\caption{ \small (\textbf{a}) Logistic map attractor. (\textbf{b}) Magnification of the box in (a).
(\textbf{c}) Noise-induced bifurcation gap in the magnified box.}
\label{Fig._13}
\end{figure}

When $\sigma >0$, the noise fluctuations smear the fine structure of the
periodic attractors as the iterate visits positions within a set of bands or
segments, like those in the chaotic attractors (see Figure \ref{Fig._13});
however, there is still a distinct transition to chaos at $\mu _{\infty
}(\sigma )$, where the Lyapunov exponent, $\lambda _{1}$, changes sign. The
period doubling of bands ends at a finite maximum period $2^{N(\sigma )}$ as 
$\mu \rightarrow \mu _{\infty }(\sigma )$ and then decreases at the other
side of the transition. This effect displays scaling features and is
referred to as the bifurcation gap \cite{schuster1, crutchfield1}.
When $\sigma$ is small, the trajectories sequentially visit the set of $%
2^{N(\sigma )}$ disjoint bands, leading to a cycle, but the behavior inside
each band is chaotic. The trajectories represent ergodic states as the
accessible positions have a fractal dimension equal to the dimension of
phase space. When $\sigma =0$, the trajectories correspond to a non-ergodic
state, since as $t\rightarrow \infty$, the positions form only a multifractal set
of fractal dimension $d_{f}=0.5338...$. Thus the removal of the noise, $%
\sigma \rightarrow 0$, leads to an ergodic to non-ergodic transition in the
map.

As shown in \cite{robglass1, robglass1b, robglass1c}, when $\mu _{\infty }(\sigma >0)$, there is
a ``crossover" or ``relaxation" time, $t_{x}=\sigma ^{r-1}$, $r\simeq 0.6332$,
between two different time evolution regimes. This crossover occurs when the
noise fluctuations begin erasing the fine structure of the attractor at $\mu _ \infty (0)$. For $t<t_{x}$, the fluctuations are smaller than the distances between the
neighboring subsequence positions of the $x_{0}=0$ orbit at $\mu _{\infty
}(0)$, and the iterate position with $\sigma >0$ falls within a small band
around the $\sigma =0$ position for that $t$. The bands for successive times
do not overlap. Time evolution follows a subsequence pattern close to that
in the noiseless case. When $t\sim t_{x}$, the width of the noise-generated
band reached at time $t_{x}=2^{N(\sigma )}$ matches the distance between
adjacent positions, and this implies a cutoff in the progress along the
position subsequences. At longer times $t>t_{x}$, the orbits no longer trace
the precise period-doubling structure of the attractor. The iterates now
follow increasingly chaotic trajectories as bands merge with time. This is
the dynamical image (observed along the time evolution for the orbits of a
single state,~$\mu _{\infty }(\sigma )$) of the static bifurcation gap
initially described in terms of the variation of the control parameter, $\mu $~\cite{crutchfield1}.

\subsubsection{Aging}

As indicated in \cite{robglass1, robglass1b, robglass1c}, the interlaced power-law position
subsequences that constitute the superstable orbit of period $2^{\infty }$
within the noiseless attractor at $\mu _{\infty }(0)$ imply a built-in aging
scaling property for the single-time function, $x_{t}$. These subsequences
are relevant for the description of trajectories that are at first ``held" at
a given attractor position for a waiting period of time $t_{w}$ and then
``released" to the normal iterative procedure. If the holding positions are
chosen to be any of those along the top band shown in the right panel of Figure \ref{trajmap} with 
$t_{w}=2k+1$, $k=0,1,...$, one obtains \cite{robglass1, robglass1b, robglass1c}:
\begin{equation}
x_{t+t_{w}}\simeq \exp _{q}(-\lambda _{q}t/t_{w})
\end{equation}
where $\lambda _{q}=\ln \alpha /\ln 2$, with $\alpha =2.50290...$. This
property is gradually removed when noise is turned on. The presence of a
bifurcation gap limits its range of validity to total times $t_{w}+t$ $%
<t_{x}(\sigma )$ and, so, progressively disappears as $\sigma$ is increased 
\cite{robglass1, robglass1b, robglass1c}.

\subsubsection{From Diffusion to Arrest}

\begin{figure} 
\begin{center}
\includegraphics[width=0.45\textwidth]{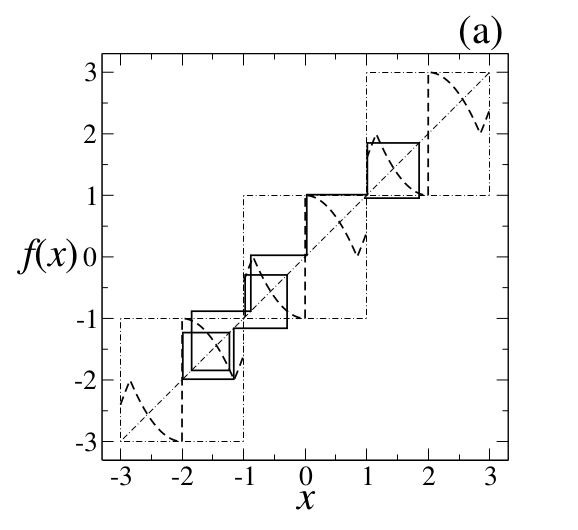} %
\includegraphics[width=0.45\textwidth]{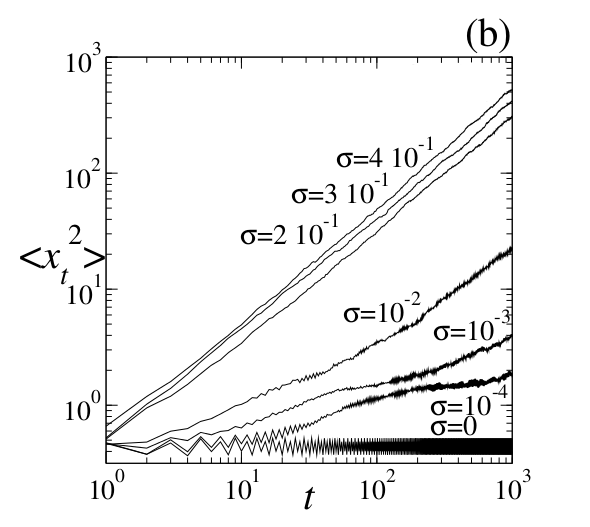}
\end{center}
\caption{ \small Glassy diffusion in the noise-perturbed onset of chaos. (\textbf{a}) Repeated-cell map (thick dashed line) and trajectory (full line). (\textbf{b}) Time evolution of the mean square displacement $\langle x_t^2 \rangle$ for an ensemble of $1000$ trajectories with initial conditions randomly distributed inside $[-1,1]$. Curves are labeled by the value of the noise amplitude. See \cite{robglass1, robglass1b, robglass1c}.}
\label{Fig:ar17}
\end{figure}

Figure \ref{Fig:ar17}a shows a repeated-cell map, $f(x)$ (together with a portion of one of its trajectories), used to investigate diffusion in the map perturbed with noise. As can be observed, the escape from the central cell into any of its neighbors occurs when $|f(x)|>1$, and this can only happen when $\sigma >0$. As $\sigma \rightarrow 0$, the escape positions are confined to values of $x$ increasingly close to $x = 0$, and for $\sigma =0$, the trajectory positions are trapped within the cell. Figure \ref{Fig:ar17}b shows the mean square displacement, $\langle x_t^2 \rangle$, obtained from an ensemble of uniformly distributed initial positions in $[-1,1]$ for several values of $\sigma$. The progression from normal diffusion to final arrest can be clearly observed as $\sigma \rightarrow 0$.

\section{Manifestations of Incipient Chaos in Complex Systems}

We turn now to the issue of the appearance of the anomalous dynamics of
critical attractors in complex systems. We present two examples. In the
first, we refer to the tangent bifurcation as the dynamical analog of rank
distributions \cite{robledo5, robzipf1b}, among which the empirical law of Zipf enjoys
a unique place, due to its combined omnipresence and simplicity. Zipf's law
refers to the (approximate) power law that is displayed by sets of data when
these are given a ranking in relation to magnitude or rate of recurrence.
The second illustration considers the application of the Horizontal
Visibility algorithm \cite{tolo1, tolo2}, which transforms time series
into networks to the anomalous dynamics at the Feigenbaum attractor \cite
{robtolo1}. In this case, we observe, in a complex network setting, the $q$-generalized
Pesin-like identity.

\subsection{A Minimal Theory for Rank Distributions}

\begin{figure} 
\centering
\includegraphics[width=0.5\textwidth]{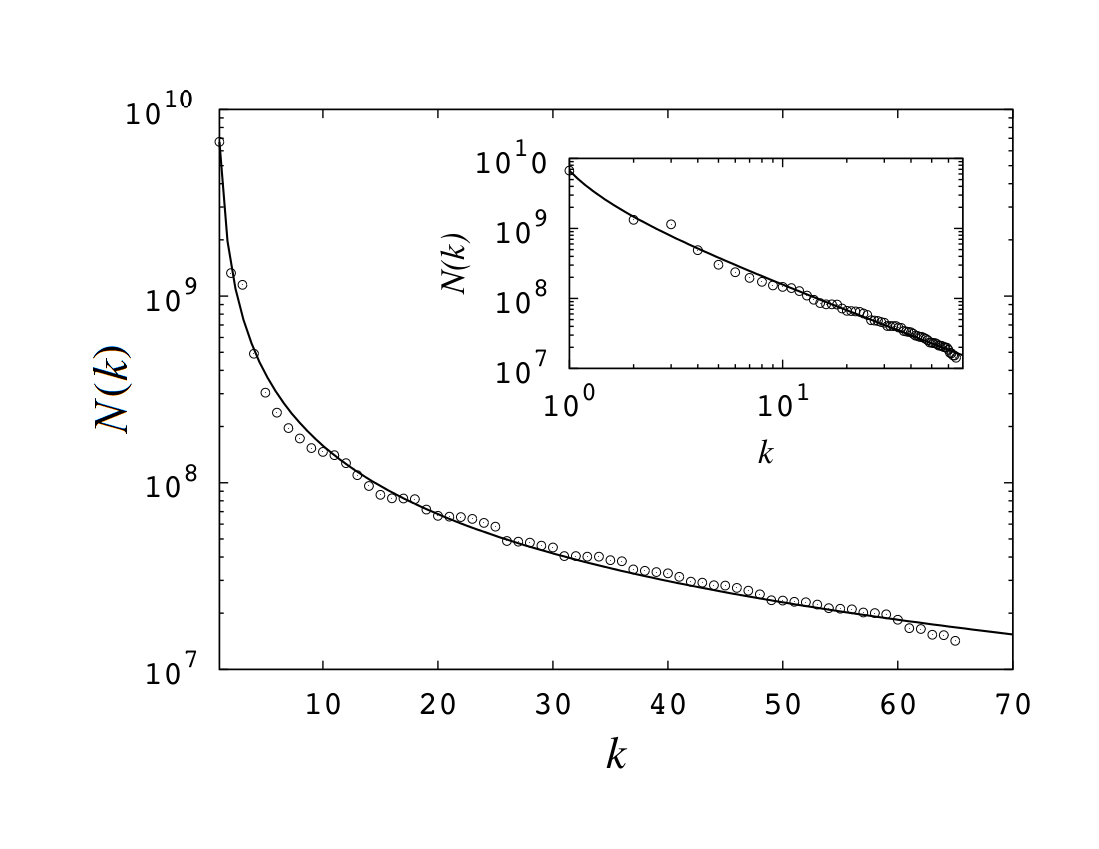} 
\caption{Rank-order statistics for the world population by country (
empty circles) taken from \textit{CIA
-The World Factbook}. The $x$-axis represents the
rank, while the $y$-axis stands for the population. Equation (\ref{zipf2}) with $\protect\alpha \simeq 1.86$ (\emph{smooth curve}) is fitted to the data.}
\label{fig:words}
\end{figure}

We start with the probability, $P(N)$, of the data, $N$, under consideration for ranking and
build the (complementary) cumulative distribution:
\begin{equation}
\Pi (N,N_{\max })=\int_{N}^{N_{\max }}P(N^{\prime })dN^{\prime }
\label{cumulative1}
\end{equation}%
where $N_{\max}$ is the maximum value in the data set of $\mathcal{N}$
numbers. Then, identify $\mathcal{N} \Pi$ with the rank $k$, that is, $\Pi
(N(k),N_{\max })=\mathcal{N}^{-1}k$. Solve Equation (\ref{cumulative1}) for $N(k)$%
, the size-rank distribution associated with $P(N)$. The (normalized)
frequency-rank distribution, $f(k^{\prime })$, is $\Pi (N,N_{\max })$ with $%
f=\Pi$ and $k^{\prime }=N$. Consider the case $P(N)\sim N^{-\alpha }$, $\alpha >1$. We obtain:

\begin{equation}
k=\mathcal{N}\int_{N(k)}^{N_{\max }}N^{-\alpha }dN=\frac{\mathcal{N}}{%
1-\alpha }\left[ N_{\max }^{1-\alpha }-N(k)^{1-\alpha }\right] ,\ \alpha
\neq 1
\label{rank1}
\end{equation}%
where $N_{\max }$ and $N(k)$ correspond, respectively, to rank $k=0$ and
nonspecific rank $k>0$. Equation~(\ref{rank1}) introduces a formal continuum space
variable for the rank $k$, and therefore, the first value of the rank is $k=0$. The consecutive values of the rank can be natural numbers by appropriately adjusting the lower limit of integration $N(k)$. In terms of the $q$-logarithmic and $q$-exponential functions (with $\alpha
=q$), Equation (\ref{rank1}) and its inverse can be written more concisely as:

\begin{equation}
\ln _{\alpha }N(k)=\ln _{\alpha }N_{\max }-\mathcal{N}^{-1}k
\label{benford2}
\end{equation}
and
\begin{equation}
N(k)=N_{\max }\exp _{\alpha }[-N_{\max }^{\alpha -1}\mathcal{N}^{-1}k]
\label{zipf2}
\end{equation}
As shown in \cite{robledo5, robzipf1b}, Equation (\ref{zipf2}) is a generalization of
Zipf's law that is capable of quantitatively reproducing the behavior for
small rank $k$ observed in real data, where, as one would expect, $N_{\max }$
is finite. In Figure~\ref{fig:words}, we compare the populations of countries
with $N(k)$ as given by Equation (\ref{zipf2}), where the reproduction of the
small-rank bend displayed by the data before the power-law behavior sets in
is evident. In the theoretical expression, this regime persists up to
infinite rank $k\rightarrow \infty $. Alternatively, we recover from Equation (%
\ref{zipf2}) Zipf's law $N(k)\sim k^{1/(1-\alpha )}$ in the limit $N_{\max
}\gg 1$ when $\alpha >1$.

\subsubsection{Rank Distributions and Their Nonlinear Dynamical Analogs}

To make explicit the analogy between the generalized law Zipf and the
nonlinear dynamics of intermittency \cite{robledo5, robzipf1b}\ , we recall the RG fixed
point map for the tangent bifurcation; see Equation (\ref{fixed1}). The
trajectories $x_t, \ t=1, 2, 3,... $, produced by this map, comply (analytically) with:%
\begin{equation}
\ln _{z}x_{t}=\ln _{z}x_{0}+ut
\label{trajectory2}
\end{equation}%
or
\begin{equation}
x_{t}=x_{0}\exp _{z}\left[ x_{0}^{z-1}ut\right]  \label{trajectory3}
\end{equation}%
where the $x_{0}$ are the initial positions. The parallel between Equations (\ref%
{trajectory2}) and (\ref{trajectory3}) with Equations (\ref{benford2}) and (\ref%
{zipf2}), respectively, is clear, and therefore, we conclude that the
dynamical system represented by the fixed-point map operates
in accordance to the same $q$-generalized statistical-mechanical properties
discussed previously. We notice that the absence of an upper bound for the
rank $k$ in Equations (\ref{benford2}) and (\ref{zipf2}) is equivalent to the
tangency condition in the map. Accordingly, to describe data with finite maximum rank, we look at the changes in $N(k)$
brought about by shifting the corresponding map from tangency (see Figure~\ref%
{fig:map}), {\it i.e.}, we consider the trajectories, $x_{t}$, with initial
positions $x_{0}$ of the map:%
\begin{equation}
x^{\prime }=x\exp _{z}(ux^{z-1})+\varepsilon ,\;0<\varepsilon \ll 1
\label{offtangency1}
\end{equation}%
with the identifications $k=t$, $\mathcal{N}^{-1}=-u$, $N(k)=x_{t}+x^{\ast }$%
, $N_{\max }=x_{0}+x^{\ast }$ and $\alpha =z$, where the translation, $%
x^{\ast}$, ensures that all $N(k)\geq 0$. In Figure~\ref{fig:eigen}, we
illustrate the capability of this approach to reproduce quantitatively real
data for the ranking of eigenfactors (a measure of the overall value) of physics
journals \cite{robledo5, robzipf1b}.

\begin{figure} 
\centering
\includegraphics[width=0.45\textwidth]{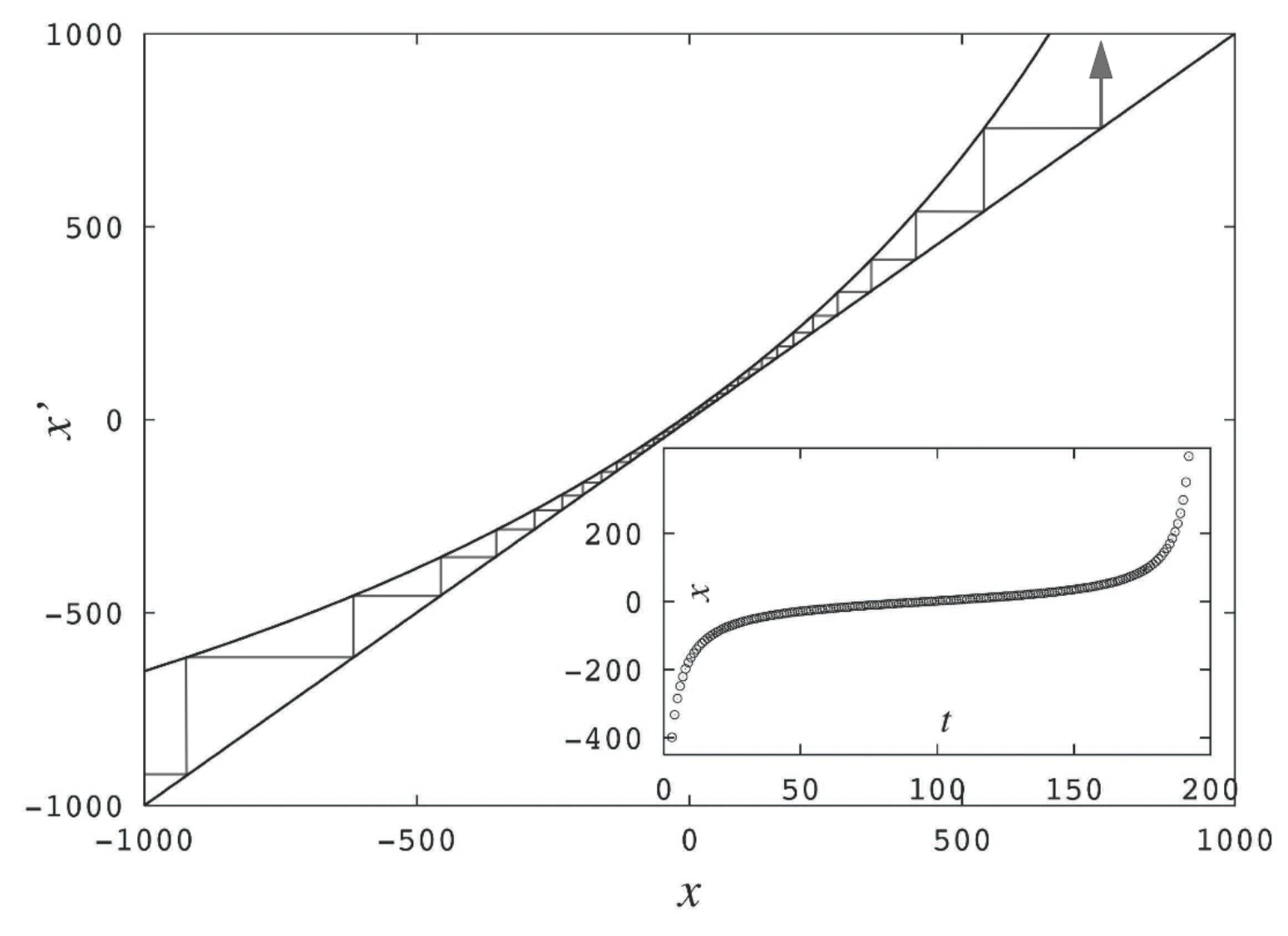}
\caption{\small The map in Equation (\ref{offtangency1}) with a trajectory. The
inset shows the time dependence of the trajectory.}
\label{fig:map}
\end{figure}

\begin{figure} 
\centering
\includegraphics[width=0.48\textwidth]{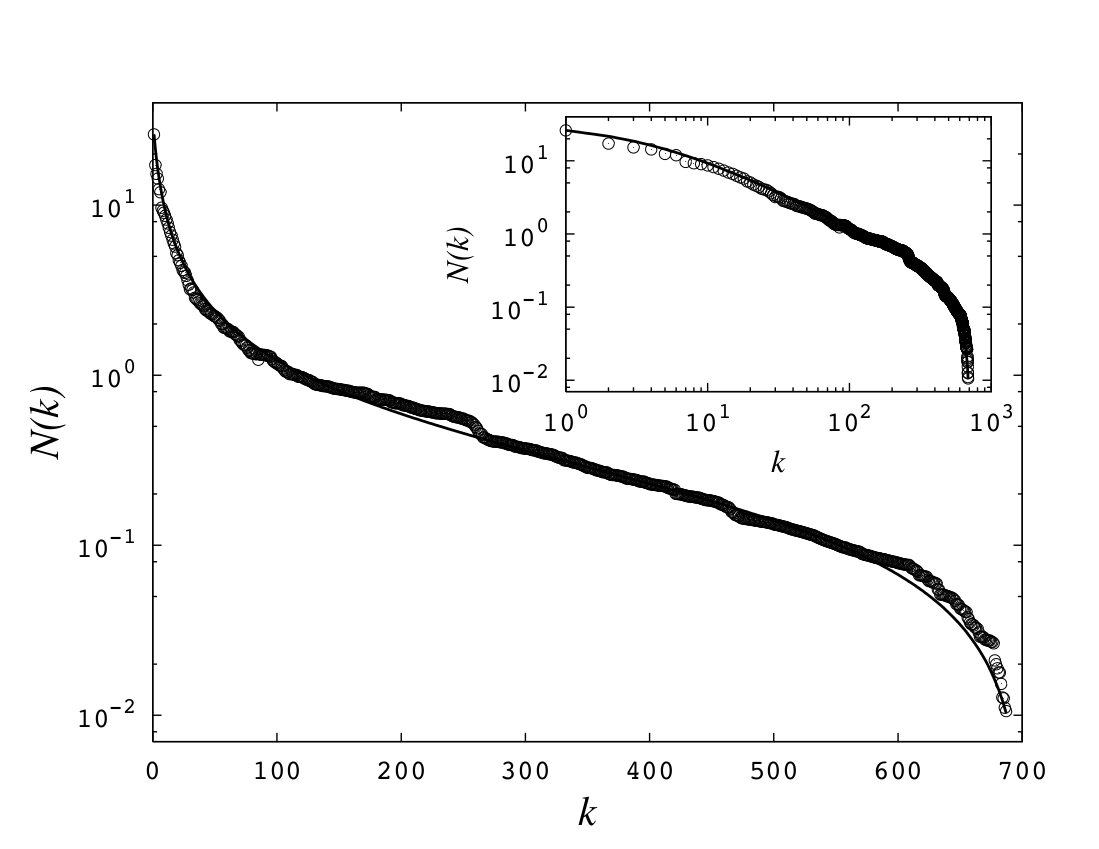}
\caption{Size-rank statistics for the eigenfactor of physics journals. Equation (\protect\ref{offtangency1}) with the
identifications provided in the text when $\protect\alpha =2.01$ and $%
\protect\epsilon =-0.00064$ (smooth curve) is fitted to the data.}
\label{fig:eigen}
\end{figure}

In the intermittency route out of chaos, it is relevant to determine the
duration of the so-called laminar episodes \cite{schuster1}, {\it i.e.}, the
average time spent by the trajectories going through the \textquotedblleft
bottleneck'' formed in the region where the map is closest to the line of the
unit slope. Naturally, the duration of the laminar episodes diverges at the
tangent bifurcation when the Lyapunov exponent for the separation of
trajectories vanishes. Interestingly, it is this property of the nonlinear
dynamics that translates into the finite-size properties of the
rank distribution function, $N(k)$. One more important result that follows from
the analogy between nonlinear dynamics and the rank law is that the most
common value for the degree of nonlinearity at tangency is $z=2$, obtained
when the map is analytic at $x=0$ with a nonzero second derivative, and this
implies $\alpha =2$, close to the values observed for most sets of real data.

\subsubsection{Rank Distributions from a Statistical-Mechanical Viewpoint}

In \cite{robledo5, robzipf1b}, it is argued that the rank distribution, $N(k)$, in Equation (\ref{zipf2}) belongs to a special type of thermodynamic structure obtained from the usual
via a scalar deformation parameter represented by the power, $\alpha$. There, Equation (\ref{benford2}) is interpreted as an
(incomplete) Legendre transform (like a Landau free energy or a free energy
density functional) between two thermodynamic potentials. The expression
relating the corresponding partition functions becomes a generalized Zipf's
law. We identify these quantities in terms of the variables involved, as well
as the conjugate variables in the transform, which are the rank $k$ and the
inverse of the total number of data, $\mathcal{N}^{-1}$. We also argued that
this kind of deformed thermodynamics arises from the existence of a strong
barrier to enter configurational phase space, which leads to only a fractal
or multifractal subset of this space being accessible to the system. A
quantitative consequence of considering $N_{\max }$ finite is the
reproduction of the small-rank bend displayed by real data before the
power-law behavior sets in. The power law regime in the theoretical
expression of $N(k)$ in Equation (\ref{zipf2}) persists up to infinite rank $k\rightarrow \infty$,
representative of a ``thermodynamic limit''.

\subsection{Complex Network Images of Time Series at the Transitions to Chaos}

The anomalous dynamics at the onset of chaos in unimodal maps has been
captured by a special kind of complex network \cite{robtolo1}. Very recently 
\cite{robtolo2, robtolo3, robtolo4, robtolo5, robtolo6}, the horizontal visibility (HV) algorithm 
\cite{tolo1, tolo2} that transforms time series into networks has
offered a view of chaos and its genesis in low-dimensional maps from an
unusual perspective that offers novel features and new understanding.
Amongst these sets of studies, we focus here on networks generated by unimodal
maps at their period-doubling accumulation points and describe briefly the
fluctuations in connectivity as the network size grows. First, it was found \cite
{robtolo1} that the expansion of connectivity fluctuations admits the
definition of a graph-theoretical Lyapunov exponent. Second, the expansion
rate of trajectories in the original critical attractor translates into a
generalized entropy that surprisingly coincides with the spectrum of
generalized graph-theoretical Lyapunov exponents \cite{robtolo1}. These are
the analogs of the $q$-entropy rates and $q$-Lyapunov exponents described earlier,
and the result suggests that Pesin-like identities valid at the onset of
chaos could be found in complex networks that possess certain scaling
properties \cite{robtolo1}. Recently, these results have been extended to the
case of the quasiperiodic route to chaos in circle maps \cite{robtolo2}.

\begin{figure} 
\begin{center}
\includegraphics[width=0.7\textwidth]{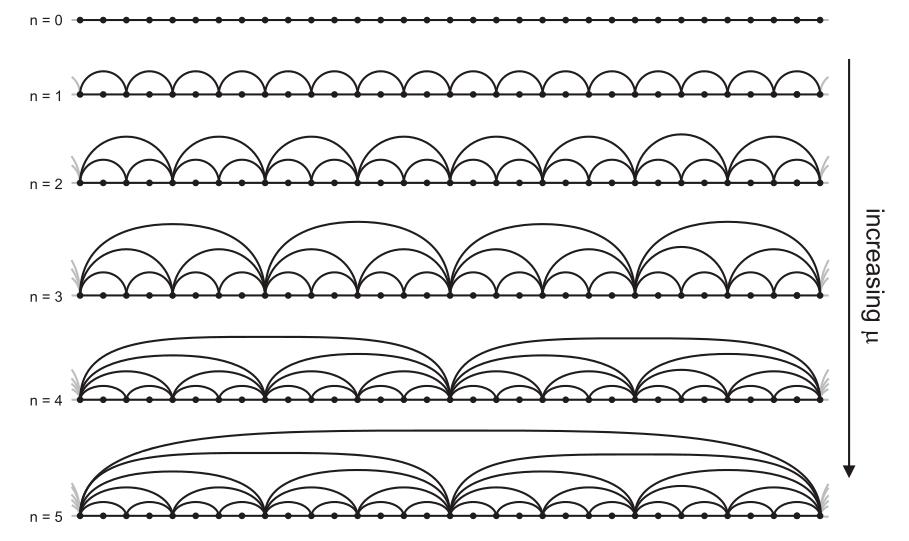}
\end{center}
\caption{ Periodic Feigenbaum graphs for $\mu < \mu_{\infty}$. The sequence of graphs associated with periodic attractors with increasing period $T = 2^n$ undergoing a period-doubling cascade. The pattern that occurs for increasing values of the period is related to the universal ordering with which an orbit visits the points of the attractor. Observe that the hierarchical self-similarity of these graphs requires that the graph for $n-1$ is a subgraph of that for $n$.}
\label{fig:feiggraphs1}
\end{figure}

\begin{figure} 
\begin{center}
\includegraphics[width=0.7\textwidth]{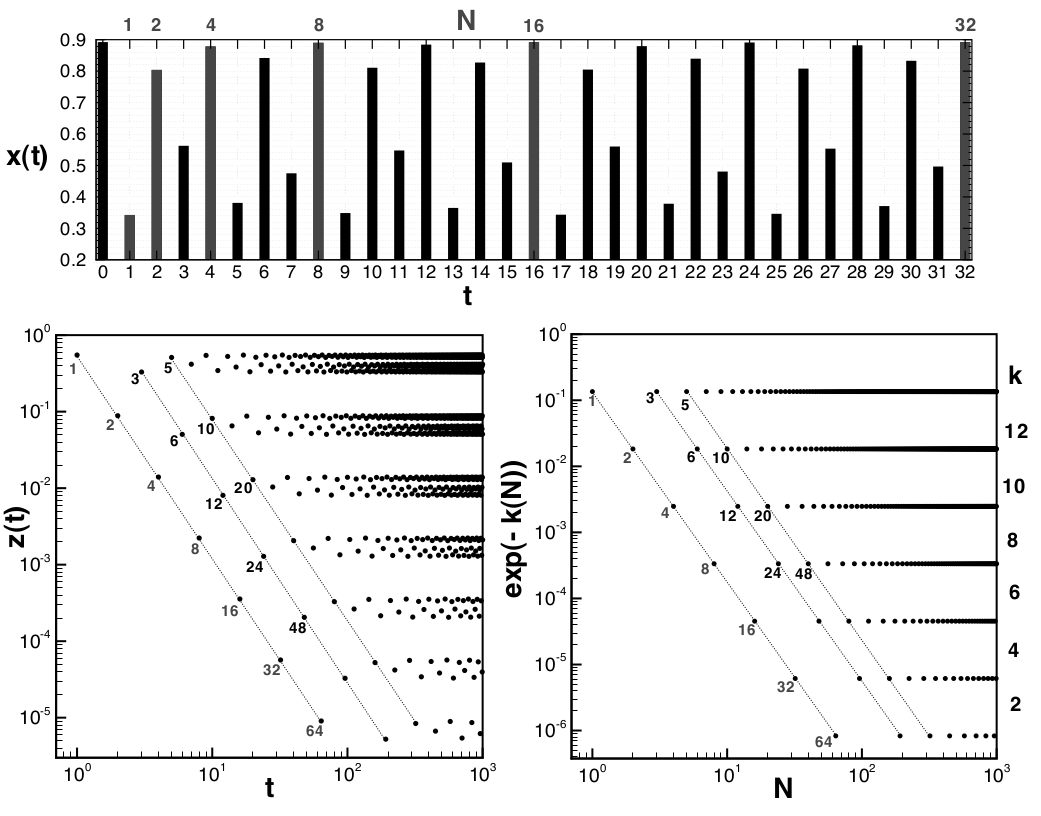}
\end{center}
\caption{ \small (\textbf{Top}) Series $x(t)$ as a function of time t for the first $10^6$ data generated from a logistic map in its version $f(x)=rx(1-x), 0 \leq x \leq1$ at the period-doubling accumulation point (only the first 33 data are
shown). The data highlighted is associated with specific subsequences of nodes. (\textbf{Left}) Log-log plot of the rescaled variable $z(t) = f (1/2) - x(t)$ as a
function of $t$, for the same series as the upper panel. This rescaling is performed to reflect the multifractal structure of the attractor. The order of visits to some specific
data subsequences is highlighted. (\textbf{Right}) Log-log plot of $\exp k( N )$ as a function of the node $N$ of the graph generated from the same time series as for the upper
panel, where $N = t$.}
\label{FigHVgraphs}
\end{figure}

The horizontal visibility (HV) algorithm is a general method to convert time
series data into a graph \cite{tolo1, tolo2} and is concisely stated as
follows: assign a node, $i$, to each datum, $x_{i}$, of the time series, $
\{x_{i}\}_{i=1,...,N}$, of $N$\ real data, and then, connect any pair of nodes, 
$i$, $j$, if their associated data fulfill the criterion, $x_{i}$, $
x_{j}>x_{n} $, for all $n$, such that $i<n<j$. The capability of the method to
transfer the most basic properties of different types of time series into their resultant
graphs has been demonstrated in recent works. See the references in \cite{robtolo2, robtolo3, robtolo4, robtolo5, robtolo6}. When the series under study are the trajectories
within the attractors generated by unimodal or circle maps, the application
of the HV algorithm yield subfamilies of visibility graphs that render the
known low-dimensional routes to chaos in a new setting \cite{robtolo1,robtolo6}. For illustrative purposes, in Figure \ref{fig:feiggraphs1},
we show a hierarchy of Feigenbaum graphs obtained along the period-doubling
bifurcation cascade of unimodal maps. In the right panel of Figure \ref{FigHVgraphs}, we show the connectivity, $k$, as a function of the node number, $N$, obtained from the trajectory at the period-doubling transition to chaos, shown in the left panel in the same figure. The HV algorithm was applied to the time series shown in the top panel in Figure \ref{FigHVgraphs}. As shown in the right panel of Figure \ref{FigHVgraphs} the distinctive band pattern of the attractor is recovered, although in a simplified manner, where the fine structure is replaced by single lines of constant degree. The order of visits to some specific node subsequences is highlighted. See \cite{robtolo1} for details.

\subsubsection{Fluctuating Dynamics and Graph-Theoretical Lyapunov Exponents}

As described in \cite{robledo8}, an alternative for capturing subexponential fluctuations
is to deform the logarithm in the definition of the ordinary Lyapunov
exponent by an amount that recovers the linear growth present for 
$\lambda_{1}>0$, that is:
\begin{equation}
\lambda _{q}=\frac{1}{t}\ln _{q}\xi _{t} \label{Lyapunov_Tsallis}
\end{equation}
where the extent of deformation $q$ is such that, while the expansion, $\xi
_{t}$, is subexponential, $\ln _{q}\xi _{t}$ grows linearly with $t$. The subexponential connectivity expansion rate for the graph under
study can be treated similarly as in \cite{robtolo1}. Since the graph is connected by construction (all nodes have degree $k\geq 2$), the variability of $k$ is only associated with a rescaled degree, $
k_{+}\equiv k-2$, we drop the subindex $+$ from now on and write $k$ meaning 
$k_{+}$. The formal network analog of the sensitivity to initial conditions $
\xi _{t}$ has as a natural definition $\xi (N)\equiv \delta (N)/\delta
(0)=\exp k(N)$, the ratio of the distance $\delta (N)=\exp k(N)$ in the
network phase space at node $N$ to the initial distance $\delta (0)=\exp
k(0)=1$, the shortest distance that the HV algorithm yields for nearby
trajectory positions \cite{robtolo1}. Accordingly, the standard network
Lyapunov exponent is defined as:
\begin{equation}
\lambda _{1}\equiv \lim_{N\rightarrow \infty }\frac{1}{N}\ln \xi (N)
\label{LyapunovN}
\end{equation}%
The value of $k(N)$ oscillates with $N$ (see Figure \ref{FigHVgraphs}), but its
bounds grow slower than $N$, as $\ln N$, and, therefore, in network context $%
\lambda _{1}=0$, in parallel to the ordinary Lyapunov exponent at the onset
of chaos. The logarithmic growth of the bounds of $\ln \xi (N)=k(N)$ is
readily seen by writing $k(N=m2^{j})=2j$ as:%
\begin{equation}
k(N)=\frac{2}{\ln 2}\ln \frac{N}{m} \label{excessdegree1}
\end{equation}
As suggested by Equation (\ref{Lyapunov_Tsallis}), we deform the the ordinary
logarithm in $\ln \xi (N)=k(N)$ into $\ln _{q}\xi (N)$ by an amount $q>1$,
such that $\ln_{q}\xi (N)$ depends linearly on $N$, and define the
associated generalized graph-theoretical Lyapunov exponent as:
\begin{equation}
\lambda _{q}=\frac{1}{N-m}\ln_{q}\xi (N) \label{Tsallis}
\end{equation}%
where we suppose that the initial time is of the
form $m\cdot 2^{0}$ (see bottom panels of Figure \ref{FigHVgraphs}, where the
expansion along the $m$ lines start always for $j=0$). One obtains:
\begin{equation}
\lambda _{q}=\frac{2}{m\ln 2}
\label{lambdaq}
\end{equation}%
with $q=1-\ln 2/2$.

\subsubsection{Entropic Functionals and Pesin-Like Identities}

To complete our argument, we remember the persistency property of trajectory
distributions of unimodal maps at the period-doubling onset of chaos. That
is, for a small interval of length $l_{0}$ with $\mathcal{N}$
uniformly-distributed initial conditions around the extremum of a unimodal
map, all trajectories behave similarly, remain uniformly-distributed at later
times and follow the concerted pattern shown in the left panel of Figure \ref{FigHVgraphs}. As
we described above (see Subsection \ref{SECdynamicsperdoubaccpoint} and Figure \ref{Fig._8}) at the period-doubling
transition to chaos, it was found that if the initial distribution is uniform
and defined around a small interval of an attractor position, the
distribution remains uniform for an extended period of time, due to the
subexponential dynamics. We denote this distribution by $\pi (t)=1/W(t)$,
where $W(t)=\mathcal{N}/l_{t}$ and $l_{t}$ is the length of the interval
that contains the trajectories at time $t$. All such trajectories give rise
to the same Feigenbaum graph \cite{robtolo1}, and at iteration times, say,
of the form $t=2^{j}$, $j=0,1,2,\dots$, the HV criterion assigns $k=2j$
links to the common node $N=2^{j}$. While $\pi$ is defined in the map, we
inquire about the $j$-dependence of $\pi (2^{j})=1/W_{j}$.

The $q$-deformed entropy:
\begin{equation}
S_{q}\left[ \pi (N)\right] =\ln _{q}W_{j}=\frac{1}{1-q}\left[ W_{j}^{1-q}-1%
\right] 
\label{q-entropy1}
\end{equation}%
where the amount of deformation, $q$, of the logarithm has the same value as
before, grows linearly with $N$, as $W_{j}$ can be rewritten as:
\begin{equation}
W_{j}=\exp _{q}[\lambda _{q}(N-m)]
\label{numberW2}
\end{equation}%
with $q=1-\ln 2/2$ and $\lambda _{q}=2/(m\ln 2)$. Therefore, if we define
the entropy growth rate:
\begin{equation}
h_{q}\left[ \pi (N)\right] \equiv \frac{1}{N-m}S_{q}\left[ \pi (N)\right]
\label{entropyrate2a}
\end{equation}%
we obtain:
\begin{equation}
h_{q}\left[ \pi (N)\right] =\lambda _{q} \label{entropyrate3}
\end{equation}

a Pesin-like identity at the onset of chaos (effectively, one identity for
each subsequence of node numbers given each by a value of $m=1,3,5,...$). See \cite{robtolo1} for details.

The fluctuations of the degree capture the core behavior of the fluctuations
of the sensitivity to initial conditions at the transition to chaos, and they
are universal for all unimodal maps. The graph-theoretical analogue of the
sensitivity grows logarithmically with the number of nodes, $N$. These
deterministic fluctuations are described by a discrete spectrum of
generalized graph-theoretical Lyapunov exponents that appear to relate to an
equivalent spectrum of generalized entropy growth rates, yielding a set of
Pesin-like identities. Therefore, the entropy expression involved is
extensive and of the Tsallis type, with a precisely defined index, $q$. A
salient feature of the application of the HV algorithm to trajectories of nonlinear dynamical systems is direct access to the degree distribution and, therefore, to the
entropy associated with it.

\section{Discussion and Conclusions}

The search and evaluation of the applicability of $q$-statistics or other
proposed generalizations of the BG canonical statistical mechanics involves
an examination of the domain of validity of the BG formalism. The
circumstances for which BG statistics fails to be applicable are believed to
be associated with situations that lack the full degree of chaotic irregular
dynamics that probes configurational phase space thoroughly, a requisite for
true equilibrium. For the critical attractors of nonlinear one-dimensional
maps, such anomalous circumstances are signaled by the vanishing of the
Lyapunov coefficient, $\lambda _{1}$, and the ergodic and mixing properties of
chaotic $\lambda _{1}>0$ trajectories are no longer present when $\lambda
_{1}=0$. At the period-doubling onset of chaos, the trajectories within the
attractor, or those that have evolved a long time towards it, are confined
to a multifractal subset of phase space with fractal dimension $d_{f}<1$.
These trajectories possess boundless memory. The dynamics at the tangent
bifurcation consists of either a monotonous evolution towards or away from
the position of tangency. It does not consider the access of trajectories to an
adjacent or neighboring chaotic region, as in the setting of \cite
{gaspard1}, or as in trajectories in conservative maps with weakly developed chaotic regions \cite{schuster1, hilborn1}. Hence, there is no
reappearance of trajectories from chaotic regions that would cause the
relaxation from the $q$-statistical regime we have found to a BG regime at
some crossover iteration time, $\tau$.

The manifestation of the anomalous dynamics at the onset of chaos in the
properties of the condensed-matter physical systems and the complex systems
we have studied implies ergodicity and/or mixing breakdown in their own nature
or circumstances. As seen in Section 5.1, the study of clusters at
criticality in thermal systems by means of the saddle-point approximation in
the LGW free energy model involves the retention of only one coarse-grained,
but dominant, configuration. This, in turn, leads to physically reasonable
cluster properties that appear to fall outside the limits of validity of the
BG theory. On the other hand, it was found that the entropy expression that
provides the property of extensivity for the estimate of the number of
cluster configurations is not the usual BG expression, but that for the $q$%
-statistics. The crossover from $S_{q}$ to that for $S_{1}$ is obtained when 
the system is taken out of criticality,
because $\delta \rightarrow 1$ makes $q\rightarrow 1$. The equivalence with 
the insulator-conductor transition in the model system described in
Section 5.2 with the transition to chaos along the intermittency route suggests
ergodicity and mixing failure at the mobility edge. Furthermore, because of the
precise analogy described in Section 6.1.1 between the dynamics at the tangent
bifurcation and the stochastic problem of rank distributions, similar implications
can be entertained for the ranking properties of these complex systems. We
have seen that the dynamics of noise-perturbed logistic maps at the chaos threshold 
exhibits the most prominent features of glassy dynamics in
supercooled liquids \cite{robglass1, robglass1b, robglass1c}. The existence of this analogy cannot be
considered accidental, since the limit of vanishing noise amplitude, $\sigma
\rightarrow 0$, involves, like in glass formation, loss of ergodicity. The occurrence of these
properties in this simple dynamical system with degrees of freedom
represented via a random noise term, and no reference to molecular
interactions, suggests a universal mechanism lying beneath the dynamics of
glass formation.

We have described the dynamical behavior at the pitchfork and tangent
bifurcations of unimodal maps of arbitrary nonlinearity $\zeta >1$. This was
accomplished via the consideration of the solution to the RG functional
composition for these types of critical attractors. Our studies have made
use of the specific form of the $\zeta $-logistic map, but the results have a
universal validity, as conveyed by the RG approach. The RG solutions are
exact and have the analytical form of $q$-exponentials; we have shown that
they are the time (iteration number) counterpart of the RG fixed-point map
expression found by Hu and Rudnick for the tangent bifurcations, and that is
applicable also to the pitchfork bifurcations \cite{baldovin1}. The $q$%
-exponential for the sensitivity, $\xi _{t}$, is exact, and we have
straightforward predictions for $q$ and $\lambda _{q}$ in terms of the
fixed-point map properties. We found that the index, $q$, is independent of $%
\zeta$ and takes one of two possible values according to whether the
transition is of the pitchfork or the tangent type. The generalized Lyapunov
exponent, $\lambda _{q}$, is simply identified with the leading expansion
coefficient, $u$, together with the starting position, $x_{0}$.

One of our most striking findings is that the dynamics at the
period-doubling accumulation point is constituted by an infinite family of
Mori's $q$-phase transitions, each associated with orbits that have common
starting and finishing positions located at specific regions of the
multifractal attractor. Each of these transitions is related to a
discontinuity in the trajectory scaling function, $\sigma (y)$, or ``diameters
ratio" function, and this, in turn, implies a $q$-exponential $\xi _{t}$ and a
spectrum of $q$-Lyapunov coefficients for each set of orbits. The
transitions come in pairs with specific conjugate indexes, $q$ and $Q=2-q$,
as these correspond to switching starting and finishing orbital positions.
Since the amplitude of the discontinuities in $\sigma$ diminishes rapidly,
in practical terms, there is only the need for the evaluation for the first few of
them. The dominant discontinuity is associated with the most crowded and
sparse regions of the attractor, and this alone provides a very reasonable
description. Thus, the special values for the Tsallis entropic index, $q$, in $%
\xi _{t}$ are equal to the special values of the variable $\mathsf{q}$ in
the formalism of Mori and colleagues at which the $q$-phase transitions take
place. We found that there is an infinite number of such special values, as there is 
an infinite number of universal discontinuities $\sigma (y)$. See \cite{robledo1} for a wider~discussion.

Then, we recall that the approach of an ensemble of trajectories to the Feigenbaum
attractor leads to a rich hierarchical structure with a generalized
statistical-mechanical expression as an emergent property, an expression
that contains an infinite number of $q$-indexes in the partition function
linked to a thermodynamic potential also associated with a $q$-deformed
exponential. Again, the values of all of these $q$-indexes are given in terms of
the universal constants contained in the trajectory function, $\sigma (y)$.
Both the trajectories within and those that approach the Feigenbaum
attractor lead to stationary distributions for their sums of visited
positions. As we have seen, these distributions capture the features of the
dynamics associated with the multifractal attractor: the families of $q$%
-exponentials in the trajectory within the attractor (Figure 3), the formation
of phase-space gaps (Figure \ref{fig:f15pre08}), the hierarchical arrangements of the repellor
positions and their preimages (Figures \ref{fig:f4pre08} and \ref{fig:f7pre08}).

In the case of the attractor-repellor fixed point that represents the
tangent bifurcation, we have seen that there is a pure $q$-statistical
regime, for the trajectories, $x_{t}$, the sensitivity, $\xi _{t}$, and the
time extensivity of the entropy, $S_{q}$. A single $q$-index occurs (and
naturally, also, its companion indexes, $2-q$ and $2-1/q$). This is reflected
in the physical systems that accept a description analogous to this
dynamics, the critical clusters, the conductance at the mobility edge and
the rank distributions. However, as we have seen, the multifractal nature of
the period-doubling accumulation point (and, similarly, the critical attractors of
the quasiperiodic route to chaos in circle maps \cite{robledo3}) requires an
infinite set of $q$-indexes and their duals, $2-q$, each, of course, given by
the infinite set of universal constants contained in the trajectory scaling
function, $\sigma (y)$. As described, there are infinite families of $q$%
-exponentials in the sensitivity within the attractor and in the
configurational weights in the partition function. The anomalous dynamics
within and towards this type of critical attractor cannot be described with
a language simpler than that required to characterize the multifractal set itself. 
However, approximations can be introduced that reduce the infinite set of $q$-indexes
to a basic pair, $q$ and $2-q$. For the Feigenbaum attractor of the quadratic
logistic map $\zeta =2$, the dominance of the two largest discontinuities in $
\sigma (y)$, $\alpha ^{-1}$ and $\alpha ^{-2}$, which correspond, respectively, to the
sparsest and tighter regions of the multifractal, suggests that a reasonable
approximation is the consideration of only these two contributions, while the
others are neglected. This approximation reduces the Feigenbaum attractor
into a two-scale multifractal and a single index parameter $q=1-\ln 2/\ln
\alpha$ \cite{robledo2, robdiaz1}. Furthermore, the time series
transformation of the trajectories at the Feigenbaum attractor into a network via the HV
algorithm reduces the multifractal nature in the dynamical properties into
a single fractal description with only a single network generalized Lyapunov
spectrum \cite{robtolo1}.

Since the early days of $q$-statistics, the presence of a duality in $q$-indexes and $q$-entropies have been noticed and documented. From a purely
operational standpoint, their source is straightforward: the inverse of a $q$-exponential or, equivalently, the inverse of the argument of a $q$-logarithm, generates the dual indexes, $q$ and $Q=2-q$. Differentiation of
these functions give the twin indexes, $q$ and $Q=1/q$, and a combination of
both yields the pair $Q=2-1/q$. Larger sets of related $q$-indexes can be
produced via iteration of these and other common algebraic and analytical
procedures. Naturally, there appear dual physical quantities associated with
them, e.g., generalized Lyapunov exponents and generalized Pesin identities.
Their physical status and meanings can be answered by considering the
results obtained for the different dynamical systems considered here. These
point out that the view that between two possible entropy expressions, only
one is to be a physical quantity, is inappropriate. We quote a statement from
an earlier study \cite{baldovin3}: ``An interesting observation about the
structure of the nonextensive formalism is that the equiprobability entropy
expression $\ln _{q}W_{t}$ can be obtained not only from $S_{q}$ in:%
\begin{equation}
S_{q}\equiv \sum_{i}p_{i}\ln _{q}\left( \frac{1}{p_{i}}\right) =\frac{%
1-\sum_{i}^{W}p_{i}^{q}}{q-1}
\label{qentropy1}
\end{equation}
but also from

\begin{equation}
S_{Q}^{\dag }\equiv -\sum_{i=1}^{W}p_{i}\ln _{_{Q}}p_{i} \label{qentropy2}
\end{equation}%
where $S_{Q}^{\dag }=S_{2-Q}=S_{q}$. The inverse property of the $q$
exponential reads $\ln _{q}(y)=-\ln _{2-q}(1/y)$ for the $q$ logarithm and
as pointed out introduces a pair of conjugate indices $Q=2-q$ with the
consequence that, while some theoretical features are equally expressed by
both $S_{q}$ and $S_{Q}^{\dag }$, some others appear only via the use of
either $S_{q}$ or $S_{Q}^{\dag }$. For instance, the canonical ensemble
maximization of $S_{Q}^{\dag }$ with the customary constraints $%
\sum_{i=1}^{W}p_{i}=1$ and $\sum_{i=1}^{W}p_{i}\varepsilon _{i}=U$, where $%
\varepsilon _{i}$ and $U$ are configurational and average energies,
respectively, leads to a $Q$-exponential weight (with $Q>1$ when $q<1$). On
the other hand, the partition function is obtained via the optimization of $%
S_{q}$. The mutual Equations (\ref{qentropy1}) and (\ref{qentropy2})
elegantly generalize the BG entropy.''

Recently \cite{thurner1, thurner2}, the two entropy expressions in
Equations (\ref{qentropy1}) and (\ref{qentropy2}) have been formally examined in
relation to the maximum entropy principle (MEP) under the assumption that
only the first three Shannon--Kinchin axioms hold. The dual entropies are
known to follow from optimization involving different constraints: one is
that quoted above, $\sum_{i=1}^{W}p_{i}\varepsilon _{i}=U$; and the other is
related to the modified ``average'' $\sum_{i=1}^{W}p_{i}^{q}\varepsilon
_{i}=U_{q}$. It is clear that the two constraints leading to the two entropy
expressions play a role in the dynamics with a single pair of dual indexes, $%
q$ and $2-q$, like in that associated with the tangent bifurcation and its
manifestations in the description of critical clusters, mobility edge and
rank distributions. However, for the case of a multifractal critical
attractor, the MEP procedure calls for some generalization in order to
accommodate the presence of an infinite set of $q$-indexes. 

The common tenet in the studies reviewed here is that the properties of the systems considered are first determined independently of any method that assumes a statistical-mechanical formalism (other than the ordinary LGW Hamiltonian in the BG formalism for critical clusters in Subsection 5.1). That is, these properties are not derived under the supposition of the applicability of generalized entropy expressions or their implications. After that, the results obtained for these systems (trajectories, sensitivity, rates of approach, \textit{etc}., and their counterparts in condensed matter and other interdisciplinarily complex systems) are analyzed in relation to generalized entropy expressions or properties derived from them. Finally, we commented on pertinent conclusions. 

\subsection*{Acknowledgments}
I am indebted to all my collaborators for their
essential contributions to the studies reported here. I am grateful to
Alvaro Diaz-Ruelas for discussions, suggestions and a careful reading of the
manuscript. Support by DGAPA
-UNAM
-IN100311 and CONACyT
-CB
-2011-167978
(Mexican Agencies) is acknowledged.




\begin{thebibliography}{99}
\bibitem{note1} 
Ergodicity is referred here in relation to the total phase space on which the map is defined.
\end{thebibliography}

\end{document}